%% file: K2-2016-BLG-0005Lb.tex
\documentclass[fleqn,usenatbib]{mnras}

\usepackage{newtxtext,newtxmath}

\usepackage[T1]{fontenc}

\DeclareRobustCommand{\VAN}[3]{#2}
\let\VANthebibliography\thebibliography
\def\thebibliography{\DeclareRobustCommand{\VAN}[3]{##3}\VANthebibliography}

\usepackage{graphicx}	

\usepackage{array}
\usepackage{graphicx,nicefrac,url}
\usepackage{wasysym}
\usepackage{makecell}
\usepackage[normalem]{ulem}

\newcolumntype{C}[1]{>{\centering\let\newline\\\arraybackslash\hspace{0pt}}m{#1}}

\title[Kepler K2C9 discovery of exoplanetary microlensing]{\emph{Kepler K2} Campaign 9: II. First space-based discovery of an exoplanet using microlensing}
\input{authors}
\date{Accepted XXX. Received YYY; in original form ZZZ}
\pubyear{2022}
\begin{document}
\label{firstpage}
\pagerange{\pageref{firstpage}--\pageref{lastpage}}
\maketitle
\begin{abstract}
We present K2-2016-BLG-0005Lb, a densely sampled, planetary binary caustic-crossing microlensing event found from a blind search of data gathered from Campaign 9 of the \emph{Kepler K2} mission (\emph{K2}C9). K2-2016-BLG-0005Lb is the first bound microlensing exoplanet \emph{discovered} from space-based data. The event has caustic entry and exit points that are resolved in the \emph{K2}C9 data, enabling the lens--source relative proper motion to be measured. We have fitted a binary microlens model to the \emph{Kepler} data, and to simultaneous observations from multiple ground-based surveys. Whilst the ground-based data only sparsely sample the binary caustic, they provide a clear detection of parallax that allows us to break completely the microlensing mass--position--velocity degeneracy and measure the planet's mass directly. We find a host mass of $0.58\pm0.04 ~{\rm M}_\odot$ and a planetary mass of $1.1\pm0.1 ~{\rm M_J}$. The system lies at a distance of $5.2\pm0.2$~kpc from Earth towards the Galactic bulge, more than twice the distance of the previous most distant planet found by \emph{Kepler}. The sky-projected separation of the planet from its host is found to be $4.2\pm0.3$~au which, for circular orbits, deprojects to a host separation $a = 4.4^{+1.9}_{-0.4}$~au and orbital period $P = 13^{+9}_{-2}$~yr. This makes K2-2016-BLG-0005Lb a close Jupiter analogue orbiting a low-mass host star. According to current planet formation models, this system is very close to the host mass threshold below which Jupiters are not expected to form. Upcoming space-based exoplanet microlensing surveys by NASA's \emph{Nancy Grace Roman Space Telescope} and, possibly, ESA's \emph{Euclid} mission, will provide demanding tests of current planet formation models.
\end{abstract}



\section{Introduction}

Microlensing remains the principal method for detecting cool, low-mass exoplanets, including planets beyond the snow-line \citep{Hwang2022,doi:10.1146/annurev-astro-081811-125518}; a demographic of particular importance for verifying theories of planetary formation. 

The core-accretion theory of planet formation predicts that gas giant planets form beyond the snow line through accretion of gas onto cores that are enlarged by the presence of solid ices. Planet-formation simulations generically predict that many massive planets forming beyond the snow line subsequently migrate inwards, a process thought to give rise to the existence of the hot Jupiter population. Simulations also indicate that lower-mass planets should exist in large numbers beyond the snow line, but that these do not typically migrate from their orbit of formation \citep[e.g.,][]{Mordasini2018,2021A&A...656A..72B}. By probing the demographics of cool, low-mass exoplanets we can therefore test planet-formation predictions directly, without the need to consider complex migration dynamics. Currently, microlensing is the only available detection method sensitive to the cool low-mass exoplanet regime. 

The statistical nature of microlensing detection means that demographic information from microlensing probes planet formation around predominately low-mass stars, the most common stellar hosts in our Galaxy. Recently, \cite{2021A&A...656A..72B} have simulated the formation of planets around low-mass stars. They find that Jupiter mass planets are not expected around hosts with mass below 0.5~M$_{\odot}$ and that Exo-Earths should be abundant for hosts above 0.3~M$_{\odot}$. These predictions are directly testable with microlensing. Indeed, the planet-host system presented in our present study is a Jupiter-mass planet orbiting a 0.6~M$_{\odot}$ host, close to the threshold predicted by \cite{2021A&A...656A..72B}. 

Microlensing sensitivity to planets with host separations at and beyond the ice line also means that it covers an important regime for planet formation. At the ice line the disk viscosity is expected to vary due to the variation in the ice fraction. This is predicted to result in a so-called migration trap close to the ice line, giving rise to a tendency for planets to pile up within this region \citep[e.g.][]{2021A&A...656A..72B}. Whilst there is some evidence for a peak in the planet-host mass ratio function, which may be indicative of such a pile up \cite[e.g.][]{2018ApJ...856L..28P}, more data is needed to test if this is in agreement or in conflict with planet formation theory.

To date, at least 133 detections involving planet-mass companions have been confirmed using microlensing\footnote{\url{https://exoplanetarchive.ipac.caltech.edu/}}, including a number that have used follow-up data from the \emph{Spitzer} space telescope \cite[e.g.][]{2021AJ....162..180Y}. However, all of these were initially flagged by ground-based observations. These include 74 by the Optical Gravitational Lens Experiment (OGLE) survey, 26 by the Microlensing Observations in Astrophysics (MOA) survey and 31 by the Korean Microlensing Telescope Network (KMTNet). Later this decade NASA's \emph{Nancy Grace Roman Space Telescope} (hereafter \emph{Roman}) will undertake a dedicated survey for exoplanetary microlensing towards the Galactic Bulge \citep{Penny2019}, whilst ESA's \emph{Euclid} mission may also undertake an exoplanet microlensing survey as an additional science activity \citep{Penny2013,2014MNRAS.445.4137M}. The \emph{Roman} Galactic Bulge Time-Domain Survey is a core community survey with a nominal goal of detecting 100 Earth-mass planets and an overall target of around 1,400 planets. Space-based microlensing surveys would also have sub-Earth mass sensitivity to planetary-mass objects that are unbound from any host, objects often referred to as free-floating planets \citep{Johnson2020}. \cite{McDonald2021} recently used data from \emph{K2} Campaign 9 of the \emph{Kepler} mission \citep[hereafter \emph{K2}C9,][]{2016PASP..128l4401H} to conduct a blind search for short timescale microlensing signals. The search revealed four new ultra-short candidate events consistent with free-floating planets of around Earth mass. These discoveries are also consistent with a previous analysis of OGLE data by \cite{2017Natur.548..183M}. The event discussed in the current paper was also found as part of this blind \emph{K2}C9 microlensing search.

\section{Microlensing theory} \label{theory}

To describe the light curve produced by a microlensing event, we can use the point-source--point-lens (PSPL) model \citep{Paczynski1986}, characterised by the time of peak magnification, $t_0$, the minimum angular impact parameter of the source to the lens normalised to the angular Einstein radius, $u_0$, and the Einstein-radius crossing time, $t_{\rm E}$, via
\begin{equation}
    A(u) = \frac{u^2 + 2}{u\sqrt{u^2 + 4}},
\end{equation}
\begin{equation}
    u(t) = \sqrt{{u_0}^2 + \bigg(\frac{t-t_0}{t_{\rm E}}\bigg)^2}.
\end{equation}
While the PSPL parameters are useful when characterising any microlensing light curve, the model fails to describe lensing systems comprised of a foreground lensing host star and a bound exoplanet, nor does it allow for the finite size of the background microlensed source star. Extracting a precise value for lens mass $M_{\rm L}$ is not possible without the presence of higher-order effects in the microlensing signal. In the absence of these there is a three-fold degeneracy between $\theta_{\rm E}$, $M_{\rm L}$ and the distance to the lens $D_{\rm L}$:
\begin{equation}\label{equation:degeneracy}
   \theta_E = \sqrt{\frac{4GM_L}{c^2}\frac{D_{\rm S} - D_{\rm L}}{D_{\rm S} D_{\rm L}}} = \sqrt{\kappa M_{\rm L} \pi_{\rm rel}},
\end{equation}
 where $D_S$ is the source distance, $\kappa = 4G/(c^2\,{\rm au})$ and  $\pi_{\rm rel}$ is the lens--source relative parallax \citep{Gould2000}.

If finite-source effects are evident on the light curve, we can begin to resolve the microlens degeneracy by fitting the normalised angular source radius, $\rho$. $\theta_{\rm E}$ can then be obtained from \citep{Gould1994,Witt1994,Nemiroff1994}.
\begin{equation}
    \theta_{\rm E} = \frac{\theta_*}{\rho},
\end{equation}
where the source star angular size $\theta_*$ can be determined via a stellar angular size versus surface brightness relation \citep[e.g.][]{1999AJ....117..521V,Yoo2004}. In addition to the lens mass and distance, the relative proper motion $\mu_{\rm rel}$ between the lens and source can be obtained from
\begin{equation}
    \mu_{\rm rel} = \frac{\theta_{\rm E}}{t_{\rm E}},
    \label{eq:murel}
\end{equation}
which can be used, along with the line-of-sight location of the lens system, to determine whether the host star is likely to reside in the Galactic disk or bulge, enabling microlensing observations to {\bf be} used statistically to explore exoplanet demographic differences between stellar populations.

\subsection{Parallax Theory}

Introducing the microlensing parallax $\pi_{\rm E}=\pi_{\rm rel}/\theta_{\rm E}$ to the model is a crucial step if the lens mass is to be directly calculated. Two methods of inducing a parallax effect into a lightcurve are via Earth-motion parallax, caused by a significant departure of the lens-source relative proper motion vector $\vec{\mu}_{\rm rel}$ from rectilinear motion and by space-based parallax, which requires observation of the same microlensing event simultaneously from two vantage points separated widely enough to result in a measurably different $u_0$ and $t_0$ between datasets \citep{Refsdal1966,Gould1992}.

The Earth-motion parallax is particularly important for well-sampled events or for those with timescales greater than 30 days \citep{Poleski2019MM}. To quantify the deviation induced in the lens-source trajectory from rectilinear motion, we must first choose a suitable reference epoch $t_{\rm 0,par}$ from which the deviation can be calculated \citep{Gould2004}. At this time, the Earth's position vector relative to the Sun and normalised to an au is $\vec{r}_\oplus(t_{\rm 0,par})=\vec{r}_{\rm \oplus,0}$ and its corresponding orbital velocity vector is $\vec{v}_{\rm \oplus,0}$. The physical deviation is thus
\begin{equation}
    \vec{\delta r}_\oplus(t) = \vec{r}_\oplus(t) - \vec{r}_{\rm \oplus,0} - (t - t_{\rm 0,par})\vec{v}_{\rm \oplus,0}.
\end{equation}
Likewise, the same logic applies to any space-based observatory, with a physical deviation denoted by $\vec{\delta r}_{\rm sat}(t)$. Using the 2-dimensional microlensing parallax vector $\vec{\pi}_{\rm E}$, which is parallel to $\vec{\mu}_{\rm rel}$ and is oriented in a celestial coordinate system with a Northern and Eastern component, $\pi_{\rm E,N}$ and $\pi_{\rm E,E}$ respectively, we can project the physical deviation into a normalised angular deviation parallel to $\vec{\mu}_{\rm rel}$, denoted by $\delta_{\rm t}$ and another transverse component denoted by $\delta_{\rm u}$, given by
\begin{equation}
    \delta_{\rm t} = \vec{\delta r} \cdot \vec{t} \pi_{\rm E} = \delta r(t)_{\rm E} \pi_{\rm E, E} + \delta r(t)_{\rm N} \pi_{\rm E, N},
\end{equation}
\begin{equation}
    \delta_{\rm u} = \vec{\delta r} \cdot \vec{u} \pi_{\rm E} = \delta r(t)_{\rm E} \pi_{\rm E, N} - \delta r(t)_{\rm N} \pi_{\rm E, E},
\end{equation}
where $\vec{t}$ and $\vec{u}$ are unit vectors parallel and perpendicular to $\vec{\mu}_{\rm rel}$, while $\delta r(t)_{\rm E}$ and $\delta r(t)_{\rm N}$ are the components of the physical deviation rotated into celestial coordinates, where the third component parallel to the vector separating the observer and source is not used. Understanding the effect of space-based parallax is then simply the result of modelling two lightcurves, one for the ground-based observatory and another for the space-based observatory, with a normalised lens-source offset $\vec{\Delta}$ between them given by
\begin{equation}
    \vec{\Delta} = \begin{pmatrix} \delta_{\rm t, sat} - \delta_{\rm t, \oplus} \\ \delta_{\rm u, sat} - \delta_{\rm u, \oplus} \end{pmatrix}.
\end{equation}
Quantifying both the ground-based and space-based parallax effect using knowledge of both Earth's and the satellite's location relative to the Sun can thus constrain the magnitude of $\pi_{\rm E}$, as well as its two components $\pi_{\rm E,N}$ and $\pi_{\rm E,E}$, allowing for a direct mass measurement assuming $\rho$ can be accurately determined.

\subsection{Binary Lensing Theory}

Note that the value acquired for $M_L$ from Equation~(\ref{equation:degeneracy}) represents the total mass of the lens system; in order to extract the mass of the planetary companion, we need to rely on binary lensing effects in the light-curve to extract the mass ratio, $q$, of the exoplanet to its host. In addition to $q$, the binary lensing model is parameterised by $s$, the angular separation of the primary and secondary lenses normalised to $\theta_{\rm E}$, and the angle $\alpha$ of the source's trajectory to the primary--secondary axis.

\begin{figure}
    \centering
    \includegraphics[width=8.5cm]{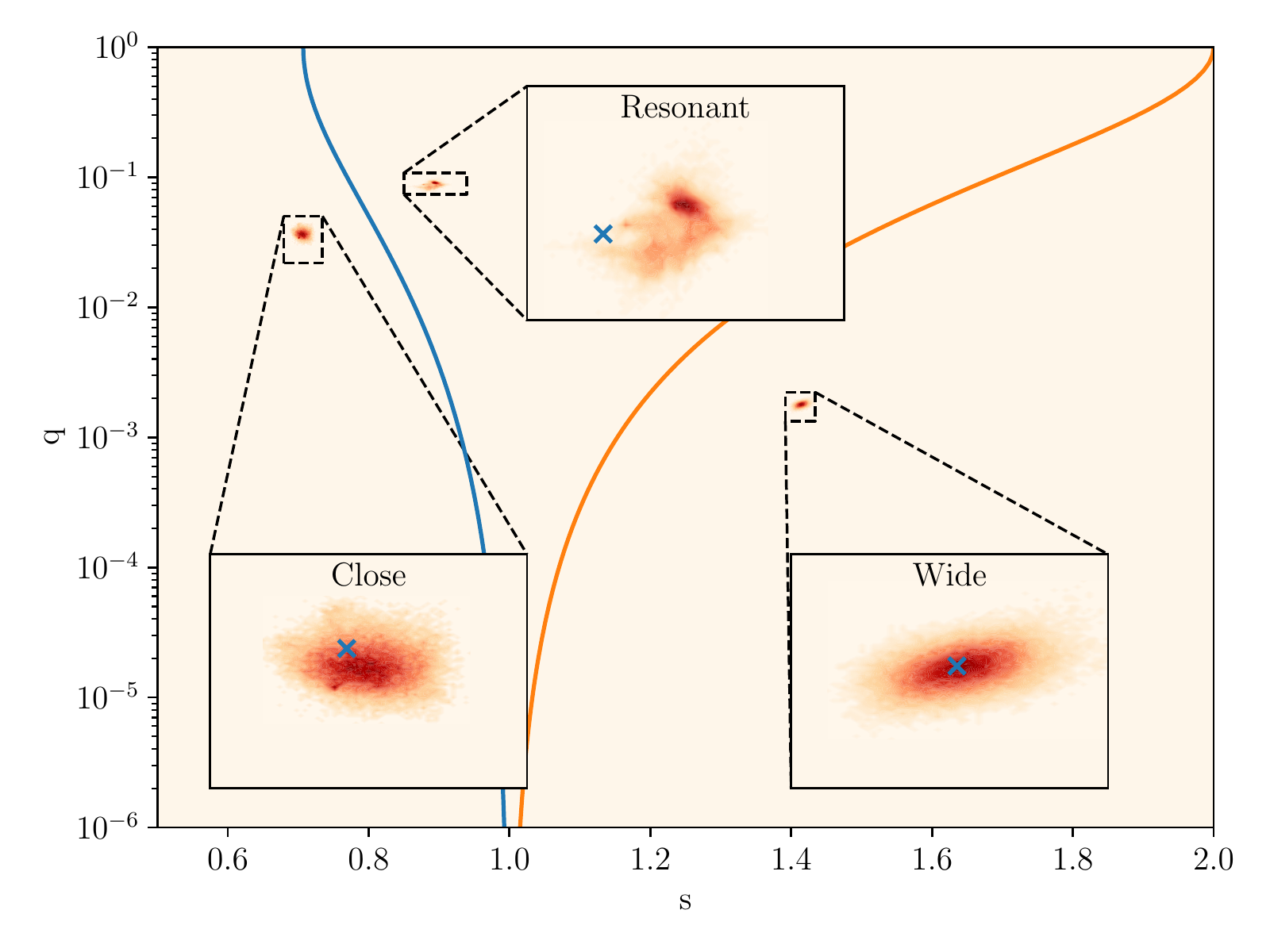}
    \caption{Shown are the distributions of the top 50\% best performing samples in $s,q$ space, from the fit for each of the three topologies investigated. The blue curve shows the boundary between the close and resonant topology, while the orange curve shows the boundary for the resonant and wide topologies. The best performing solution for each fit is indicated with a blue cross. Overall, our analysis shows the wide model to be most strongly preferred by the data, which is backed up by the proximity of the best fit wide solution to the centroid of the samples.}
    \label{fig:topology}
\end{figure}

Binary lensing effects manifest themselves in the light curve as deviations from the PSPL model through caustic crossings and cusp approaches. Caustics are locations in the source plane where a point source would be theoretically infinitely magnified upon crossing. In the case of binary lensing, these caustics have three different possible topologies, namely the close, resonant and wide topologies (see Figure~\ref{fig:topology}), with the particular topology defined entirely by the parameters $s$ and $q$ \citep{Erdl1993}. Caustic crossings result in sharp increases in source flux and, when resolved by observations, they allow strong constraints to be placed on the possible combinations of $s$ and $q$. Cusps, on the other hand, are sharp corners on the caustic shape that project regions of high magnification outward from the center of the caustic. Their influence on the light curve is typically evident when the source trajectory passes near them producing, under certain source trajectories, a secondary symmetrical peak that can further constrain $s$ and $q$.

Due to the often small size of planetary caustics compared to $\theta_{\rm E}$, an alternate binary `planetary' parameterisation to the standard $q$, $s$ and $\alpha$ can be used. The parameters $t_{0,\rm pl}$, $t_{\rm E,\rm pl}$ and $u_{0,\rm pl}$ characterise the time of closest approach of the source to the planetary caustic, the timescale of the planetary lens, and the minimum impact parameter of the source to the planetary caustic, respectively, normalised to the Einstein  radius of the planet \citep[see][for a detailed description of a similar parameterization]{Penny2014}. We found that the fitting was much more sensitive to this reparameterisation, helping the fitting to converge much more efficiently to viable solutions. The conversion from this planetary parameterisation to the conventional binary parameterisation is given in Appendix~\ref{section:appendix}.

Without other higher-order effects, such as parallax or cusp approaches, the binary lensing model can suffer from a degeneracy between the close and wide caustic topologies, requiring a thorough investigation of both models. The close--wide degeneracy is studied in this work for K2-2016-BLG-0005Lb where we show that, in this case, it is convincingly broken.

\section{Observations and Photometric Reduction} \label{sect:obs} 

\subsection{\emph{Kepler K2} Campaign 9}

\begin{figure}
    \centering
    \includegraphics[width=8.5cm]{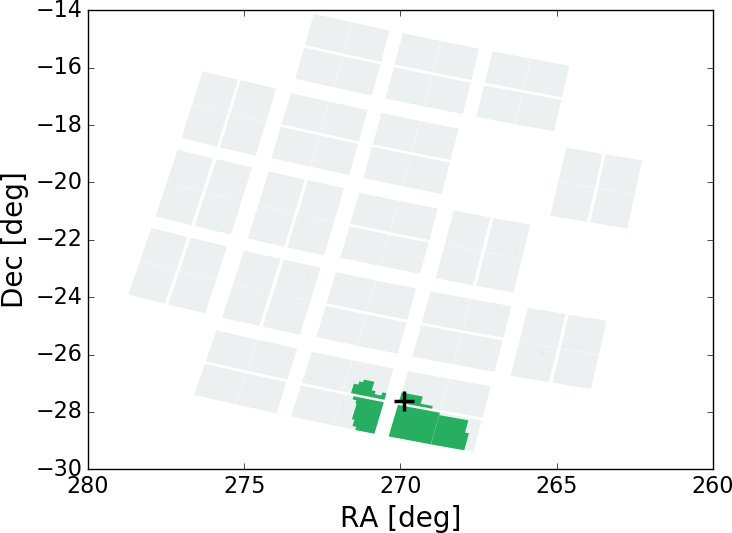}
    \caption{Kepler focal plane footprint (gray) with the K2C9 ``superstamp'' region in green. Data for K2C9 was collected only from inside this superstamp region due to data bandwidth limitations. K2-2016-BLG-0005 is located on Module 10.2 of the \emph{Kepler} focal plane array, and its position is shown by the black cross. This plot was generated using the K2C9 Visibility Tool: \url{http://k2c9.herokuapp.com/}.}
    \label{fig:k2c9Field}
\end{figure}

\emph{K2}C9 surveyed a $\sim$3.7 deg$^2$ region of the Galactic Bulge continuously with 30~min observing cadence between 22~April and 19~May 2016 (subcampaign C9a), and again from 22~May to 1~July 2016 (subcampaign C9b). \emph{Kepler} is in an Earth trailing orbit with a period of 372.5 days. At the time of K2C9 \emph{Kepler} was approximately 0.8~au from Earth. The selected sub-region of the \emph{Kepler} field, known as the ``superstamp'' (Figure~\ref{fig:k2c9Field}), was predicted to exhibit a large number of microlensing events \citep[e.g.,][]{2016A&A...595A..53B}. The data were blindly searched by \citet{McDonald2021} for short-timescale microlensing events, indicative of free-floating planets. They found five new candidate microlensing events, together with 22 additional events that had been previously catalogued by ground-based survey teams.

One of the five new events, K2-2016-BLG-0005, is a clear binary-lens event. During subcampaign C9a the event exhibits two sharp peaks, characteristic of a source crossing of microlensing caustics generated by the interaction of two lenses. There is also a marked difference between the light curve generated from the \emph{K2} photometry and data for the same event obtained simultaneously by ground-based surveys.

\begin{figure*}
    \centering
    \includegraphics[width=18cm]{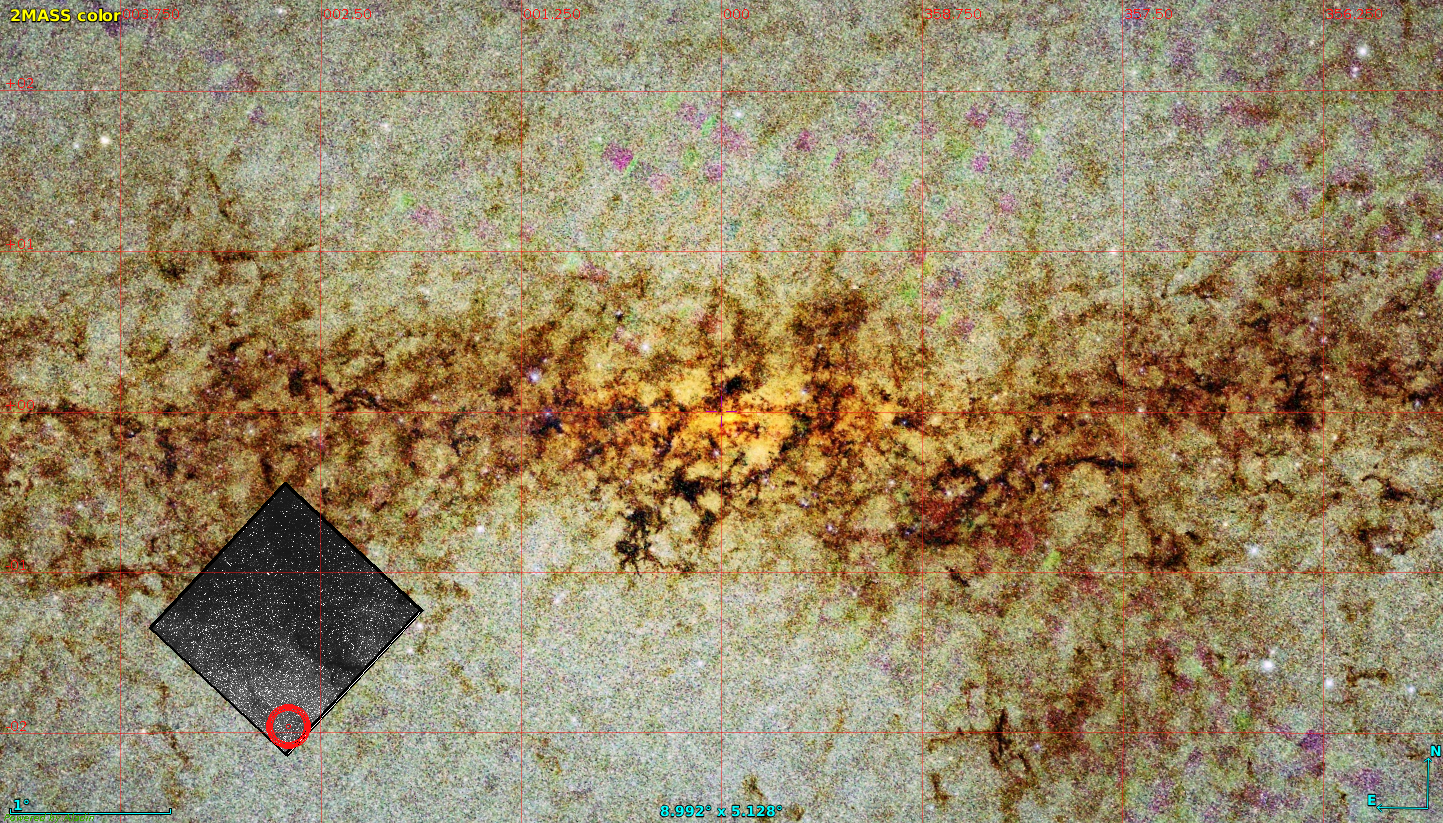}
    \caption{2MASS view of the Galactic Centre with the location and orientation of Module 10.2 of the \emph{Kepler} focal plane array shown inset. Galactic North is upwards in this image and Galactic East is to the left. The approximate location of K2-2016-BLG-0005 on the module is shown by the red circle. }
    \label{fig:2mass}
\end{figure*}

K2-2016-BLG-0005 is located at equatorial coordinates RA(J2000)~$= 17\mbox{h}59\mbox{m}31.16\mbox{s}$ Dec(J2000)~$=-27^{\circ}36'26''.90$. Figure~\ref{fig:2mass} shows the location of the event on Module 10.2 of the \emph{Kepler} focal plane, as well as the position and orientation of the module with respect to the Galactic Centre.

Calibrated \emph{K2}C9 image data were obtained from the Mikulski  Archive  for  Space  Telescopes\footnote{\url{https://archive.stsci.edu/k2/}}. A detailed description of the photometric reduction and candidate selection that led to the discovery of this event is given in \citet{McDonald2021}. 

\subsection{Ground-based observations}\label{sect:ground}

\emph{K2}C9 gave a unique opportunity to measure microlensing parallaxes for free-floating planets, hence, a wide-ranging ground-based observing campaign was organized \citep{2016PASP..128l4401H}. Higher cadence observations of the \emph{K2}C9 superstamp region were obtained by the three main microlensing surveys: KMTNet \citep[three telescopes at different sites]{Kim16}, MOA \citep{Bond01,Sako08}, and OGLE \citep{Udalski15}. All three surveys publish alerts on ongoing microlensing events on a daily basis \citep{Bond01,Udalski03,Kim18a} and KMTNet has also presented their data for \emph{K2}C9 \citep{Kim18b}. 

In addition to data from the main survey teams, we have also used data from  K2C9-CFHT Multi-Color Microlensing Survey \citep{Zang18} that was organised specifically for the K2C9 survey effort. Additionally, near-infrared $H$-band photometry has been obtained from the UKIRT Microlensing Survey \citep{2017AJ....153...61S}. 

Despite this intensive ground-based campaign, none of the ground-based surveys flagged K2-2016-BLG-0005 in advance of the blind K2C9 data search by \cite{McDonald2021}. Only after this study presented this event did the survey teams extract their data for this event.

A summary of the ground-based data is provided in Table~\ref{table:telescopes}, including the number of data points that fall within the binary caustic anomaly. Due to the abundance of ground-based data obtained before and after the K2C9 observing period, the long timescale behaviour of the event due to the microlensing effect from the host star lens is very well characterised. Ground-based photometry was extracted using dedicated implementations of Difference Image Analysis \citep[DIA;][for CFHT, KMTNet and UKIRT, MOA, OGLE respectively]{Alard00,Albrow09,Bond01,Udalski08}. The photometric uncertainties estimated by DIA are known to be underestimated and we corrected for this underestimation 
by multiplying the original uncertainties by a factor specific to a given dataset -- see second last column in Table~\ref{table:telescopes}. For the OGLE data this factor was taken from \citet{Skowron16} and for other data we determine this factor following \citet{Yee12}.

\begin{table*}
\caption{Summary of ground-based datasets.}
\label{table:telescopes}
{
\begin{tabular}{ lrrrrrrr }
    \hline\hline
    Telescope/Field & Diameter & Camera & Pixel scale & Filter & Number & \textbf{Uncertainty} & Median \\
     & (m) & field of view & (${\rm arcsec}/{\rm pix}$) & & of epochs & \textbf{scaling} &uncertainties\\
      & & (deg.$^2$) & & & $7511-7518$ & \textbf{factor} & (mag)  \\
    \hline \vspace{-8pt}\\
    CFHT Maunakea& $3.58$ & $0.94$ & $0.187$ & $g$ & 9 & $1.7$ & 0.274 \\
     & & &  & $i$ & 10 & $1.7$ & 0.113  \\
     & & &  & $r$ & 10 & $1.7$ & 0.162  \\
    KMT Aus./BLG03 & $1.6$ & $4.0$ & $0.40$ & $I$ & 78 & $1.62529$ & 0.210 \\
    KMT Aus./BLG43 & & & & $I$ & 81 & $1.56986$ & 0.174 \\
    KMT Chile/BLG03 & $1.6$ & $4.0$ & $0.40$ & $I$ & 40 & $1.5887$ & 0.216 \\
    KMT Chile/BLG43 & & & & $I$ & 40 & $1.3327$ & 0.196 \\
    KMT S.Africa/BLG03 & $1.6$ & $4.0$ & $0.40$ & $I$ & 72 & $1.45611$ & 0.195 \\
    KMT S.Africa/BLG43 & & & & $I$ & 58 & $1.38723$ & 0.178 \\
    MOA Mt. John & $1.8$ & $2.2$ & $0.58$ & $MOA_R$ & 84 & $1.49255$ & 0.528 \\
    OGLE Las Campanas & $1.3$ & $1.4$ & $0.26$ & $I$ & 45 & $1.7267^a$ & 0.116\\
    UKIRT Maunakea & 3.8 & 0.19 & 0.40 & $H$ & 11 & $2.32665$ & 0.247  \\
    \hline
 \hline
\end{tabular}
}
\newline
\footnotesize $^a$ -- $0.0029~\mathrm{mag}$ was added in quadrature following \citet{Skowron16}, but has negligible impact in this case.

\end{table*}

\subsection{\emph{K2} photometric reduction}\label{sect:photom}

\begin{figure*}
    \centering
    \includegraphics[width=18cm]{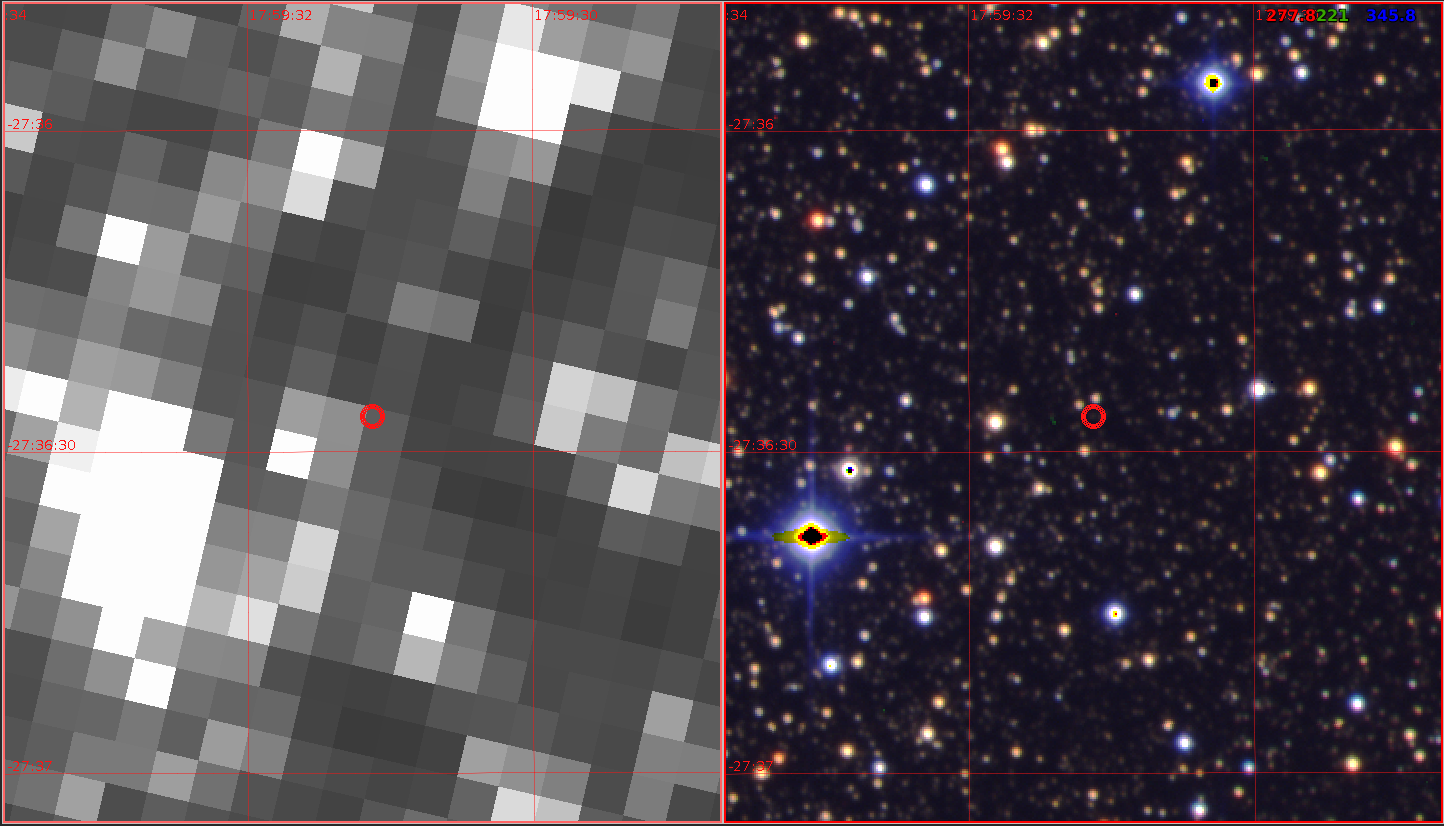}
    \caption{Comparison of K2C9 and CFHT images centered at the position of K2-2016-BLG-0005 (red circle). \emph{Left:} A small section of a K2C9 full-frame image captured during subcampaign C9a. The pixel scale is $3\farcs 98$. The event is not visible on this frame, which was taken 7 days before the start of the binary caustic anomaly. \emph{Right:} A CFHT MegaCam sloan $i+r+g$ colour composite image of the same region with a pixel scale of $0\farcs 187$. In both images Celestial North points upwards and East is to the left.}
    \label{fig:k2c9-cfht}
\end{figure*}

\begin{figure}
    \centering
    \includegraphics[width=8.5cm]{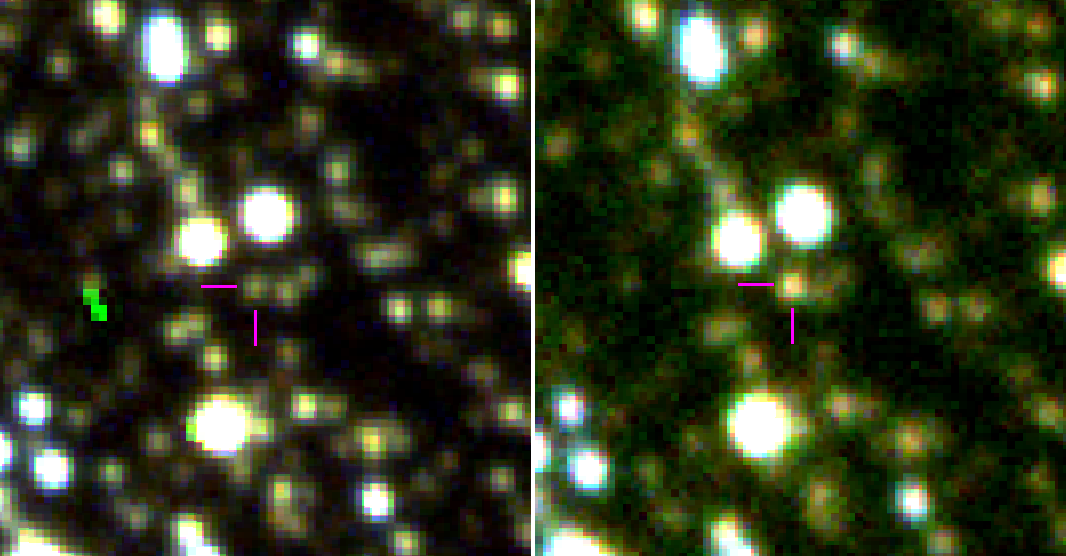}
    \caption{Colour images from CFHT showing the field around K2-2016-BLG-0005 outside (left) and inside (right) of the caustic crossing. Celestial North points upwards and East to the left. The magenta cross-hair locates the microlensed source.}
    \label{fig:colourImage}
\end{figure}

One of the principal challenges for obtaining reliable microlensing photometry from \emph{K2} stems from the relatively large \emph{Kepler} pixel size of $3\farcs 98$ compared to the dense stellar crowding towards the inner Galactic bulge, as shown in Figure~\ref{fig:k2c9-cfht}. The large pixel size means that the stellar point-spread function is heavily under-sampled, whilst the dense stellar crowding means that microlensing variations are significantly diluted by the presence of non-microlensed starlight within the pixel of the microlensed source. Another major challenge comes from the degraded pointing stability of \emph{Kepler} during the \emph{K2} mission, owing to the loss of two out of four reaction wheels, as described in \citet{Howell2014}. This gives rise to drifting of stars across detector pixels with resulting intra-pixel photometric variation induced by the non-uniform pixel profile. This is a very substantial effect for which standard photometry pipelines are ill-equipped. Whilst the ground-based data are handled using standard DIA pipelines our \emph{K2}C9 photometric reduction and fitting employ the Modified CPM photometric method \cite[{\tt MCPM},][]{2019AA...627A..54P} that was designed specifically for \emph{K2}C9 data processing. The temporal density and quality of K2 photometry after processing with {\tt MCPM} is shown in Figure~\ref{fig:wide_space}. The extraction of photometry ignores five epochs that significantly differed from other epochs ($BJD-2450000$ of $7502.97646$, $7514.42039$, $7508.33058$, $7511.72289$, and $7516.42308$). The uncertainties of the \emph{K2}C9 photometry are assumed to be $2.4$ times larger than the estimates that are based on noise information attached to the \emph{K2} images. The final list of 100 pixels used for training is attached as on-line supporting information.

\section{Microlens model fitting}

Due to the close--wide degeneracy in binary microlensing models, three different binary lens model topologies were investigated. The wide topology involves a crossing of the wide planetary caustic, while the close topology itself has a degeneracy between a single caustic approach (hereafter referred to as close topology) and a double caustic approach (hereafter referred to as resonant topology).

We sampled the posterior distributions of parameters using the Markov Chain Monte Carlo method implemented in the \texttt{EMCEE} package \citep{Foreman-Mackey13}. For calculating trajectories as well as source and blending fluxes we used the \texttt{MulensModel} package \citep[][version 2.7.2]{Poleski2019MM}. For microlensing parallax calculations one needs the time-series of positions of Earth and \emph{Kepler}, which \texttt{MulensModel} evaluates using the ERFA library\footnote{\url{https://zenodo.org/record/3564896}} and the JPL/Horizons system\footnote{\url{https://ssd.jpl.nasa.gov/horizons/}}, respectively.
The magnification of the binary lens was evaluated using \texttt{VBBL} \citep{2018MNRAS.479.5157B} with the \citet{Skowron12} polynomial root solver. \emph{K2}C9 data were extracted using \texttt{MCPM}, which decomposes the noise seen in target pixels (caused by spacecraft motion and high stellar density) into a linear combination of signals observed in other pixels. The scaling factors of this linear combination are regularized in order to prevent over-fitting. The calculation of the optimal linear coefficients that best isolate the microlensing signal requires a prior model for the signal (in this case a binary microlens model). Since the final form of the binary microlensing model is initially unknown, and is itself being fitted for, the fitting process involves an iterative scheme. The model parameters are updated via \texttt{EMCEE}, which in turn updates the photometry of the light curve via \texttt{MCPM}. This iterative joint data--model fitting process is all handled within \texttt{MulensModel}. A further important constraint on the behaviour of the fitting process ultimately comes from joint fitting with the ground-based data, which does not require \texttt{MCPM} reduction but is nonetheless coupled to the \emph{Kepler} photometry through the microlensing model.

\begin{figure*}
    \centering
    \includegraphics[width=\textwidth]{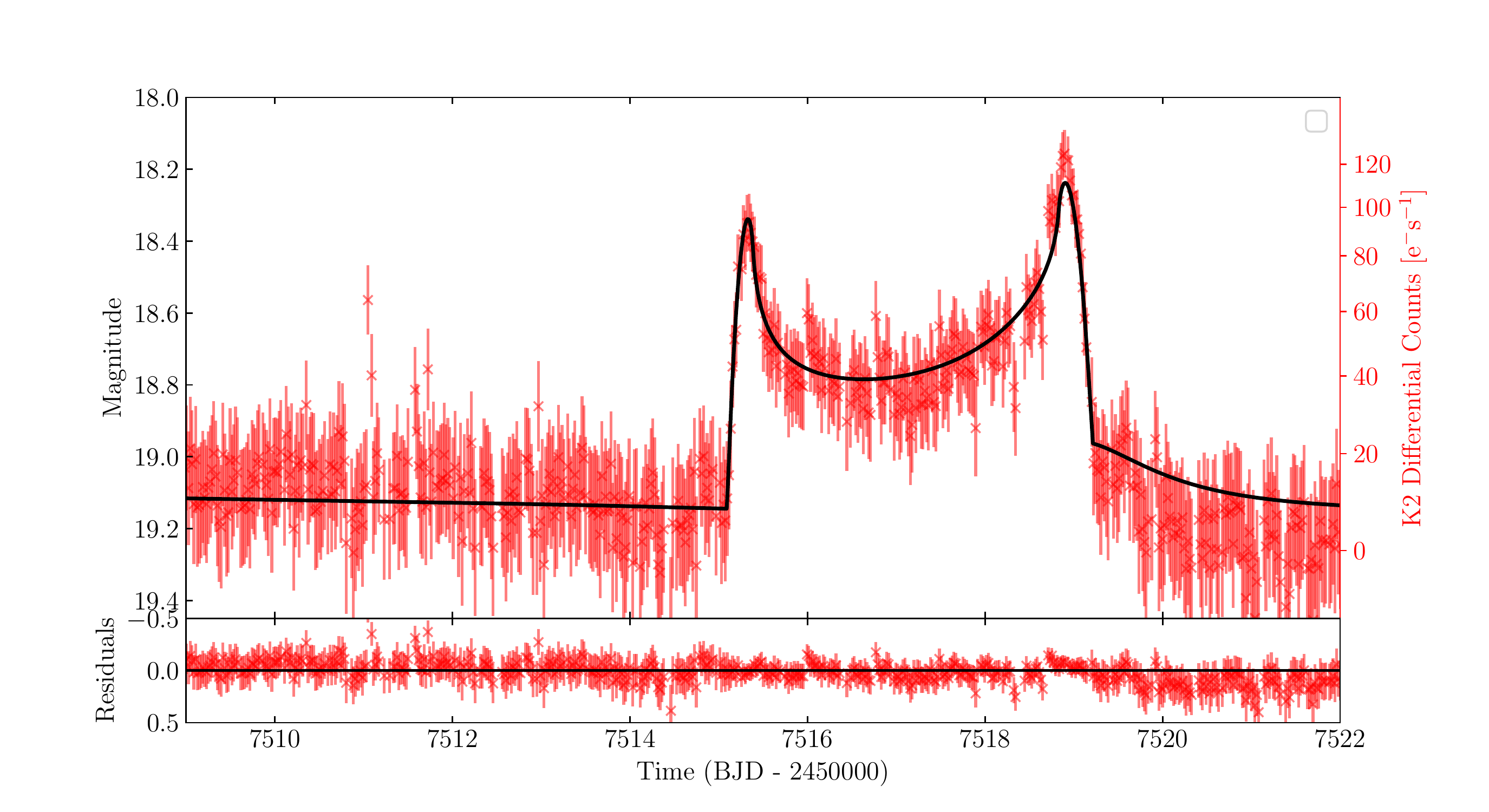}
    \caption{K2 \texttt{MCPM} photometry of K2-2016-BLG-0005Lb (red). The space-based component of the best-fit wide-topology, parallax binary lens model is shown in black. The caustic crossing region is clearly visible and well sampled between $BJD-2450000=7515$ and $7519$. The \emph{K2} differential flux counts are indicated on the right vertical axis; note that this scale is neither linear, nor logarithmic.}
    \label{fig:wide_space}
\end{figure*}
The model fitting was applied in five stages for each topology. In each stage, the number of \texttt{EMCEE} steps and walkers used in the fit was chosen to allow fitting in a reasonable amount of time, while also producing well-mixed chains and parameter distributions that were well described by Gaussian functions. The first stage involved fitting a PSPL model to the ground-based data, excluding the binary lens signatures centred around $BJD-2450000 = 7517$, to model the long-timescale behaviour due the host lens. For this stage we used \texttt{EMCEE} with 4,000 steps and 20 walkers. The walkers had wide starting distributions of the $u_0$, $t_0$, and $t_{\rm E}$ parameters. The second stage used the parameters from the initial PSPL fit as starting parameters to fit a binary-lens model with finite-source effects (and without microlensing parallax) to the ground data only, providing $\rho$, $t_{0,\rm pl}$, $t_{\rm E,\rm pl}$ and $u_{0,\rm pl}$. In this stage we used 3,000 iterations and 30 walkers. Using a $t_{0,\rm pl}$ centred on the caustic crossing, a $u_{0,\rm pl}$ of zero and a $t_{\rm E,\rm pl}$ comparable to the width of the caustic crossing gave a starting set of binary parameters when performing the fit. The third stage used \texttt{MCPM} to extract the \emph{K2} photometry for the event, starting from the best binary-lens model found in the second stage as the starting parameters and distributions. An example of iterated K2C9 {\tt MCPM} photometry (after all modeling stages) is shown in Figure~\ref{fig:wide_space} for the wide topology. These first three modeling stages enabled the final two stages to fit a joint ground--space parallax model.

\begin{table*}
\caption{Summary of model parameters for the wide caustic topology, showing results from both the regular wide fit and the ecliptic degenerate solution. The planetary binary parameterisation has been converted to the conventional $s$, $q$ and $\alpha$ formalism.}
\label{table:table0}
\centering
\resizebox{\textwidth}{!}{\begin{tabular}{ cccccccccccc }
    \hline\hline
    Model & \thead{$t_0$\\- 2450000\\(BJD)}  & $u_0$ & $t_E$ (days) & $\rho$ & $s$ & $q$ & $\alpha$ (degrees) & $\pi_{{\rm E},N}$ & $\pi_{{\rm E},E}$ & $f_{s_{sat}}$\\ 
    \hline \vspace{-8pt}\\
    Wide & $7486.6\pm0.9$ & $0.620\pm0.008$ & $76\pm2.1$ & $0.00187\pm0.00007$ & $1.414\pm0.007$  & $0.0018\pm0.0001$ & $302.3\pm4.6$ & $-0.110\pm0.003$ & $-0.0450\pm0.0017$ & $19.7\pm0.5$\\[2pt] 
    Wide Ecl. & $7488.0\pm0.8$ & $-0.629\pm0.008$ & $74\pm2.1$ & $0.00193\pm0.00007$ & $1.417\pm0.007$  & $0.0019\pm0.0001$ & $58.4\pm4.7$ & $0.090\pm0.002$ & $-0.076\pm0.0026$ & $21.5\pm0.4$\\[2pt] 
    \hline
 \hline
\end{tabular}}
\end{table*}

\begin{figure*}
    \centering
    \includegraphics[width=\textwidth]{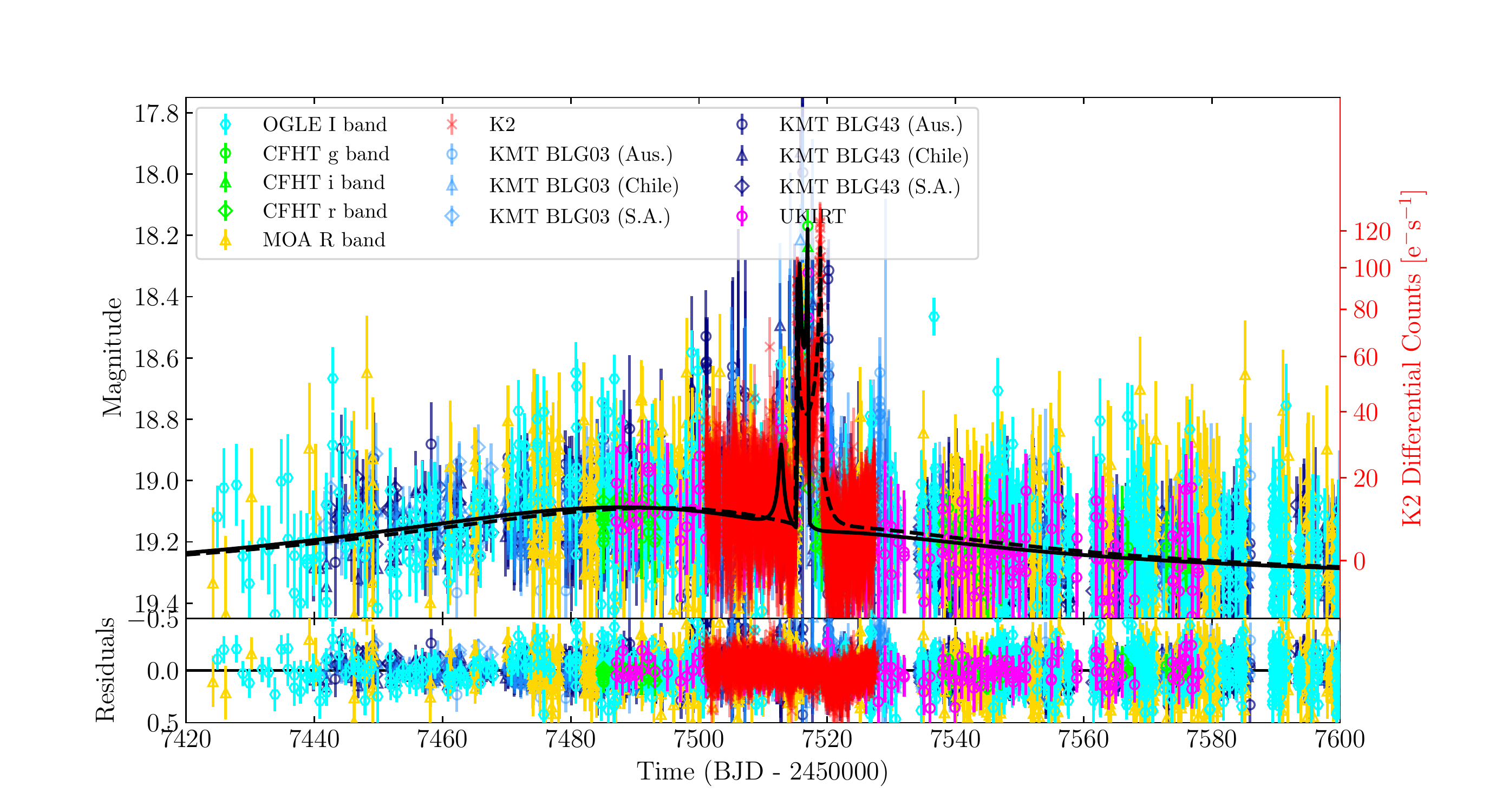}
    \caption{An expanded view of the wide model fit, showing the full range of ground-based photometry and the longer timescale magnification effect of the host star lens.}
    \label{fig:photometry}
\end{figure*}

\begin{table*}
\caption{The lens properties, extracted using the CFHT $g$ calibration, showing results from both the regular wide fit and the ecliptic degenerate solution.}
\label{table:lens_prop}
\centering
\begin{tabular}{ cccccc }
    \hline\hline
    Model & $M_{\rm L}$ (M$_\odot$) & Planet mass (M$_J$) & $D_{\rm L}$ (kpc) & $\mu_{\rm rel}$ (mas year$^{-1}$) & Projected separation (au)\\
    \hline
    Wide & $0.584 \pm 0.038$ & $1.10 \pm 0.09$ & $5.20 \pm 0.24$ & $2.71 \pm 0.07$ & $4.16 \pm 0.32$\\
    Wide (Ecliptic) & $0.574 \pm 0.037$ & $1.16 \pm 0.09$ & $5.26 \pm 0.25$ & $2.73 \pm 0.08$ & $4.11 \pm 0.32$\\
    \hline
 \hline
\end{tabular}
\end{table*}
\begin{figure*}
    \centering
    \includegraphics[width=\textwidth]{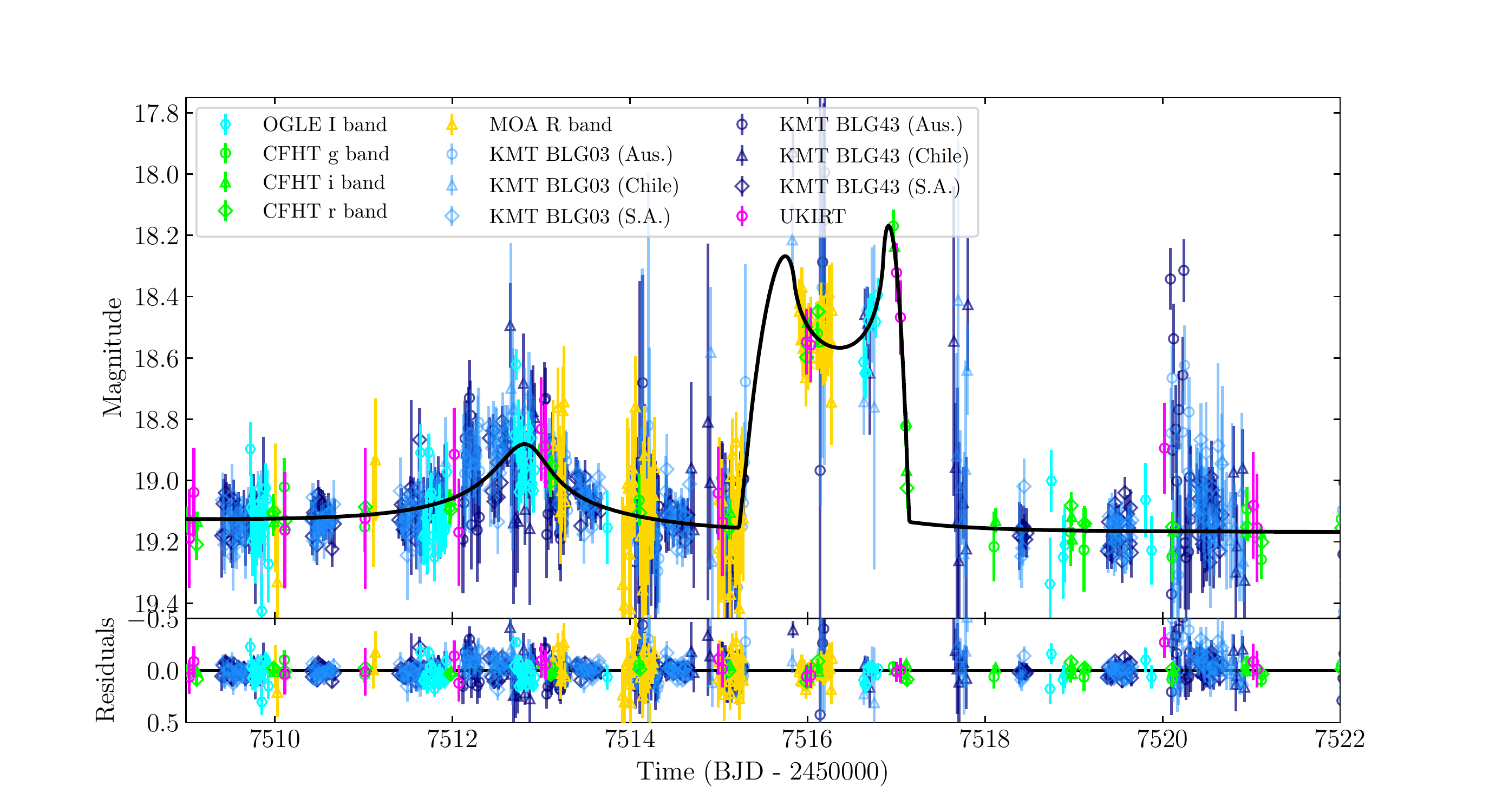}
    \caption{Ground-based photometry of K2-2016-BLG-0005Lb from KMTNet, MOA, OGLE, and CFHT datasets. The ground-based component of the best-fit wide-topology, parallax binary lens model is shown in black. This is the same lensing model as used in Figure \ref{fig:wide_space} for \emph{K2} photometry. Here, the cusp approach is evident around $BJD-2450000=7513$, with a narrow caustic crossing shown between $7515$ and $7517$. Note how the caustic exit is captured in both CFHT and UKIRT data.}
    \label{fig:wide_ground}
\end{figure*}
The difference in timing of the second caustic crossing between the ground photometry and the \emph{K2} photometry is evident in Figures~\ref{fig:wide_ground} and \ref{fig:wide} and is around one day. The observed duration between caustic crossings is also around 2.5 times longer as seen from \emph{Kepler}'s location than from the ground. This contrast highlights the validity of using a parallax model to fit to this event. In addition to the benefits of obtaining a parallax measurement, both ground-based photometry and space-based photometry provide their own advantages. Due to the high cadence of the \emph{K2} data, the caustic crossings are thoroughly sampled, while the ground-based data span multiple $t_{\rm E}$, allowing for the accurate fitting of the PSPL parameters. Both datasets are shown superposed in Figure~\ref{fig:wide}. An expanded view that shows the full extent of ground-based photometry covering the host-lens microlensing signal is provided in Figure~\ref{fig:photometry}.

\begin{figure*}
    \centering
    \includegraphics[width=\textwidth]{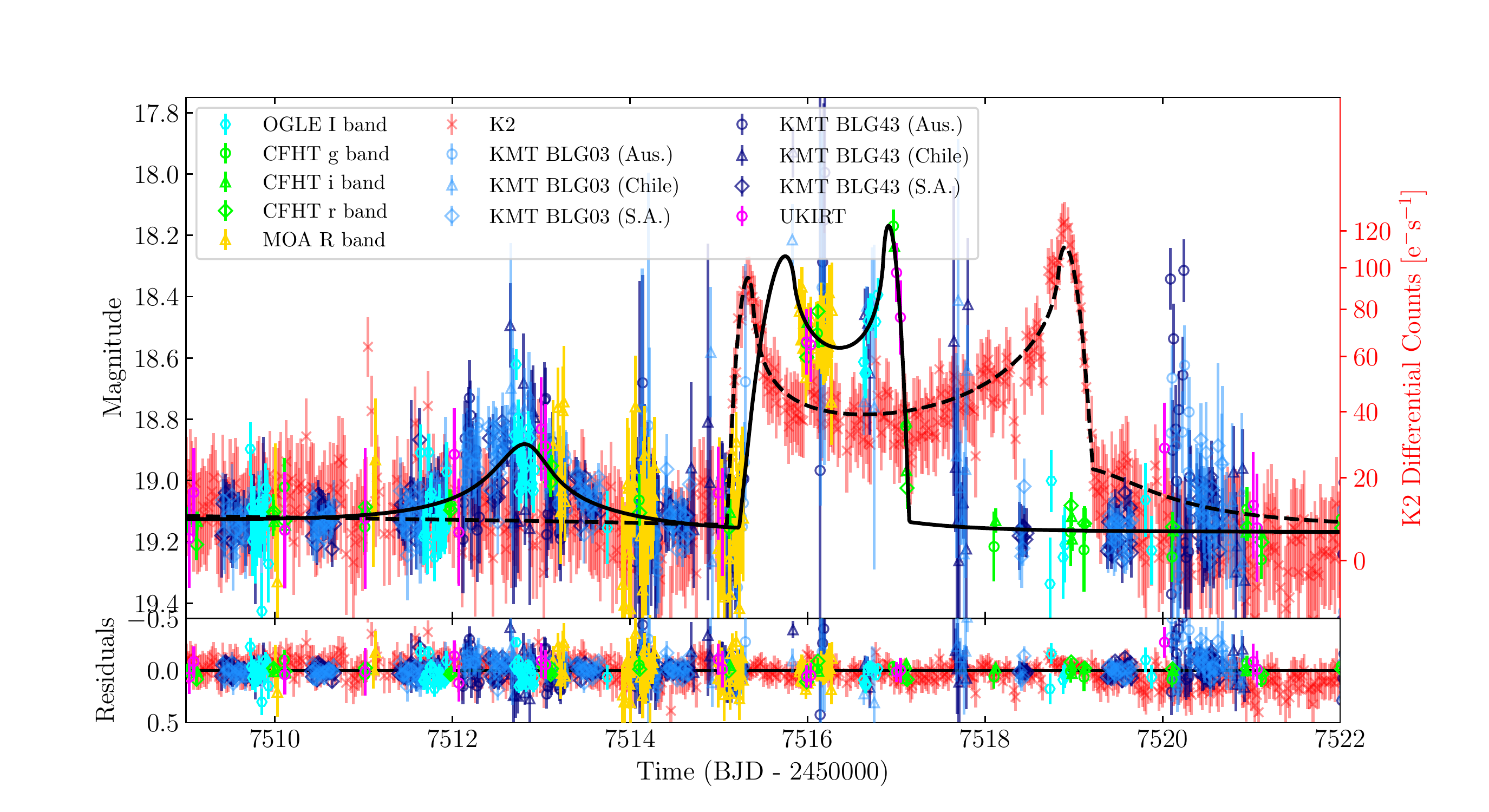}
    \caption{The superposition of photometry and best-fit model from Figures~\ref{fig:wide_space} and \ref{fig:wide_ground}. The \emph{K2} differential flux count scale is shown on the right vertical axis. The space-based component of the wide-topology solution is indicated by the dashed black line, while the ground-based component is shown as a solid black line. The best-fit wide solution provides good characterisation of the \emph{K2}C9 caustic structure, as well as ground based coverage of the caustic exit and of the pre-caustic peak seen around BJD$-2450000 = 7512.5$~days.}
    \label{fig:wide}
\end{figure*}

As introducing a space-based parallax adds the parallax components $\pi_{{\rm E},N}$ and $\pi_{{\rm E},E}$ on top of the seven parameters from the first two stages, the fourth stage used a lower iteration count of 1,000 but an increased number of walkers (100). The best model from the fourth stage was then used as the starting point for the final stage, which used 4,000 iterations with 20 walkers. Best fit parameters are given in Table~\ref{table:table0}, where the binary parameterisation has been converted back into the more familiar $q$, $s$, and $\alpha$ form (see Appendix~\ref{section:appendix}).

After fitting parallax models for each of the close, resonant and wide topologies, the wide caustic model provided a significantly superior fit, with $\chi^2_{\rm wide} = 10,141$ (reduced $\chi^2 = 0.87$). The best performing close model gave $\chi^2_{\rm close} = 10,834$ (reduced $\chi^2 = 0.93$, $\Delta \chi^2 = \chi^2_{\rm close} - \chi^2_{\rm wide} = 693$) and the best resonant model reached $\chi^2_{\rm resonant} = 11,493$ (reduced $\chi^2 = 0.99$, $\Delta \chi^2 = \chi^2_{\rm resonant} - \chi^2_{\rm wide} = 1,352$). Whilst the reduced-$\chi^2$ values seem reasonable for all three models, this is only the case because most of the ground-based data samples the long-timescale behaviour of the host lens and therefore the reduced-$\chi^2$ statistic is not strongly sensitive to behaviour of the data during the anomaly. The $\Delta \chi^2$ statistic provides a clearer view of the relative support for each model and shows that the wide model is significantly more favoured by the data than either of the other models. The wide topology solution is shown in Figures~\ref{fig:wide_space}-\ref{fig:photometry}, whilst the best-fit close and resonant solutions are shown in Figures~\ref{fig:close} and \ref{fig:resonant}, respectively, in Appendix~\ref{section:appendix}.

\begin{figure*}
    \centering
    \includegraphics[width=14cm]{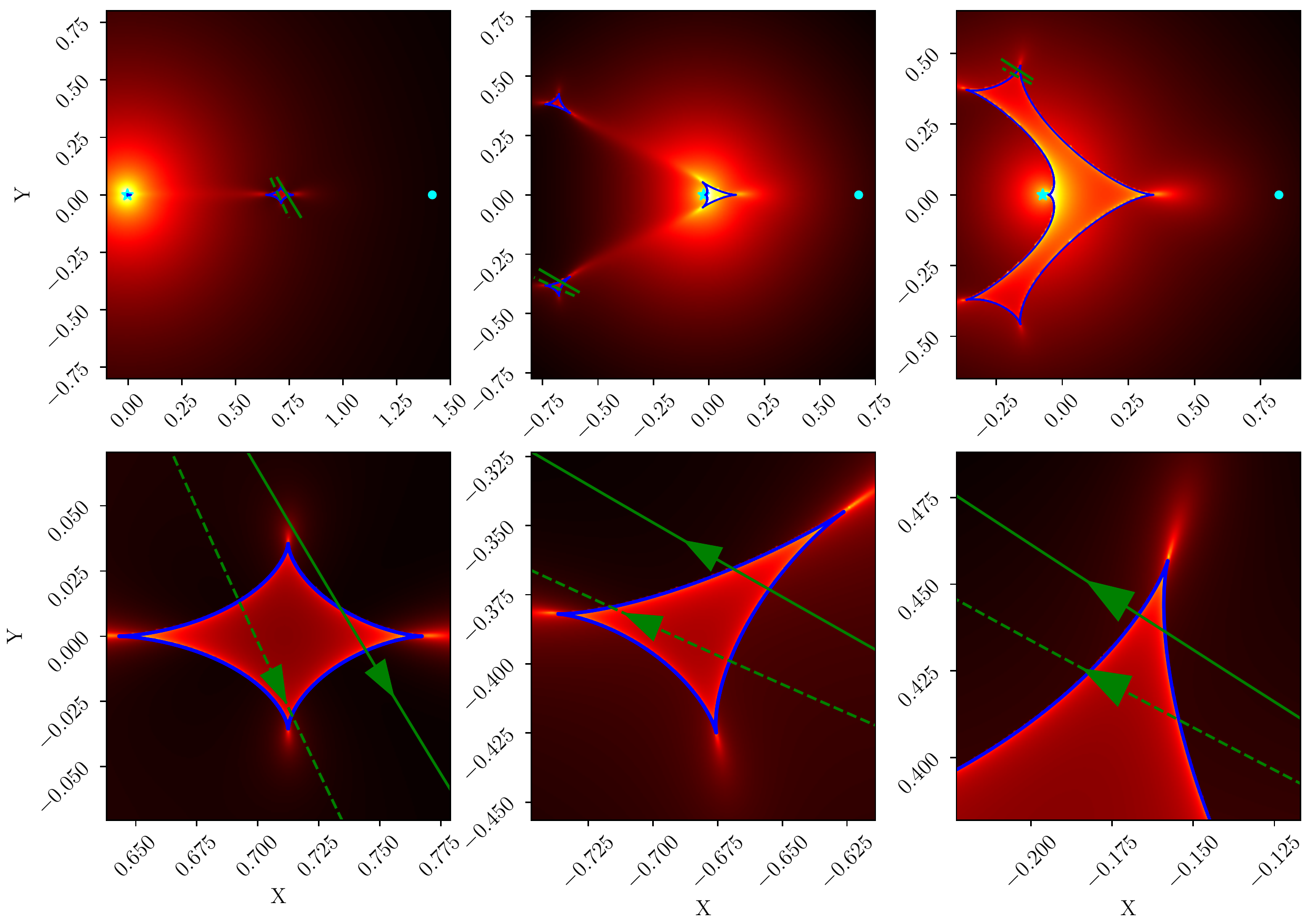}
    \caption{The trajectories of the source from the \emph{K2} (green dotted line) and ground (green solid line)  based vantage points are shown superimposed onto the caustics (blue lines) for the wide (left), close (middle) and resonant (right) topologies, with the bottom row showing a zoomed in version of the caustic crossing. The magnification maps are shown underneath on a logarithmic scale to illustrate the spatial variation of the cusps and caustic shapes. For figures on the top row, the location of the primary (host) lens is indicated with a cyan star, while the secondary (planetary) lens is indicated with a cyan circle, further to the right. The separation of the lenses is given by the binary parameter $s$, with the centre of mass located at the origin.}
    \label{fig:caustics}
\end{figure*}
The wide solution has the advantage that the ground-based trajectory passes near a caustic cusp around ${\rm BJD}-2450000 = 7512.5$ in Figure~\ref{fig:wide}, matching the ground photometry well, which has coverage over the duration of the approach. By contrast, the close and resonant solutions do not allow for a cusp approach at that epoch. In the case of the resonant topology, the best fit solution involves a significantly larger mass ratio ($s=0.89$ and $q=0.09$) than for the close or wide solutions but is the least favoured of the three as it also provides a poor description of the caustic behaviour seen in the \emph{K2}C9 data. The data overall therefore strongly favour a planetary model for this event. The positioning of the cusps relative to the caustic crossings for each topology and fit is shown by the source trajectories in Figure~\ref{fig:caustics}.

The full corner plot for the wide model showing all of the parameter covariances and the marginalised parameter distributions is shown in Figure~\ref{fig:cornerTop} of Appendix~\ref{section:appendix}. The distribution of each fitting parameter is well characterised by a Gaussian. Strong correlations are shown between $t_{\rm E}$ and $t_0$, $u_0$ and $t_{0,\rm pl}$, $u_0$ and $u_{0,\rm pl}$ and between $t_{\rm E}$ and $\pi_{{\rm E},N}$, which exist to ensure the epoch of the caustic crossing is respected.

Since the Galactic bulge is located close to the ecliptic plane, parallax events detected towards the bulge are subject to a two-fold degeneracy known as ecliptic degeneracy \citep{2005ApJ...633..914P}. This degeneracy is exact for events located on the ecliptic and can result in two different valid solutions for the lens mass arising from different velocity solutions. K2-2016-BLG-0005 is located at ecliptic latitude $\beta = -4\fdg 16$ so we should expect the wide model to exhibit near-degenerate solutions. For this reason we ran a further fit for the wide model to determine the second solution. The trial parameters for the ecliptic degenerate solution involve a simple transformation $u_0 \rightarrow -u_0$, $\alpha \rightarrow -\alpha$ and $\pi_{\rm E,N} \rightarrow -\pi_{\rm E,N}$ \citep{2005ApJ...633..914P}. The resulting fit for the second wide solution is shown in Figure~\ref{fig:ecl_sol} and the fit parameters for both solutions are given in Table~\ref{table:lens_prop}. The ecliptic degenerate wide solution has $\chi^2_{\rm wide,2} = 10,162$, which is very similar to that of the first solution ($\Delta \chi^2 = 21$). So, as expected, both degenerate models provide a good fit to the data, with neither having convincing statistical support over the other. However, as can be seen from Table~\ref{table:lens_prop}, their best-fit parameters are consistent within error so, whilst the data cannot distinguish between either model, the resulting lens parameters are essentially unaffected by the degeneracy.

\subsection{Source Characterization}

\begin{figure}
\includegraphics[width=8cm]{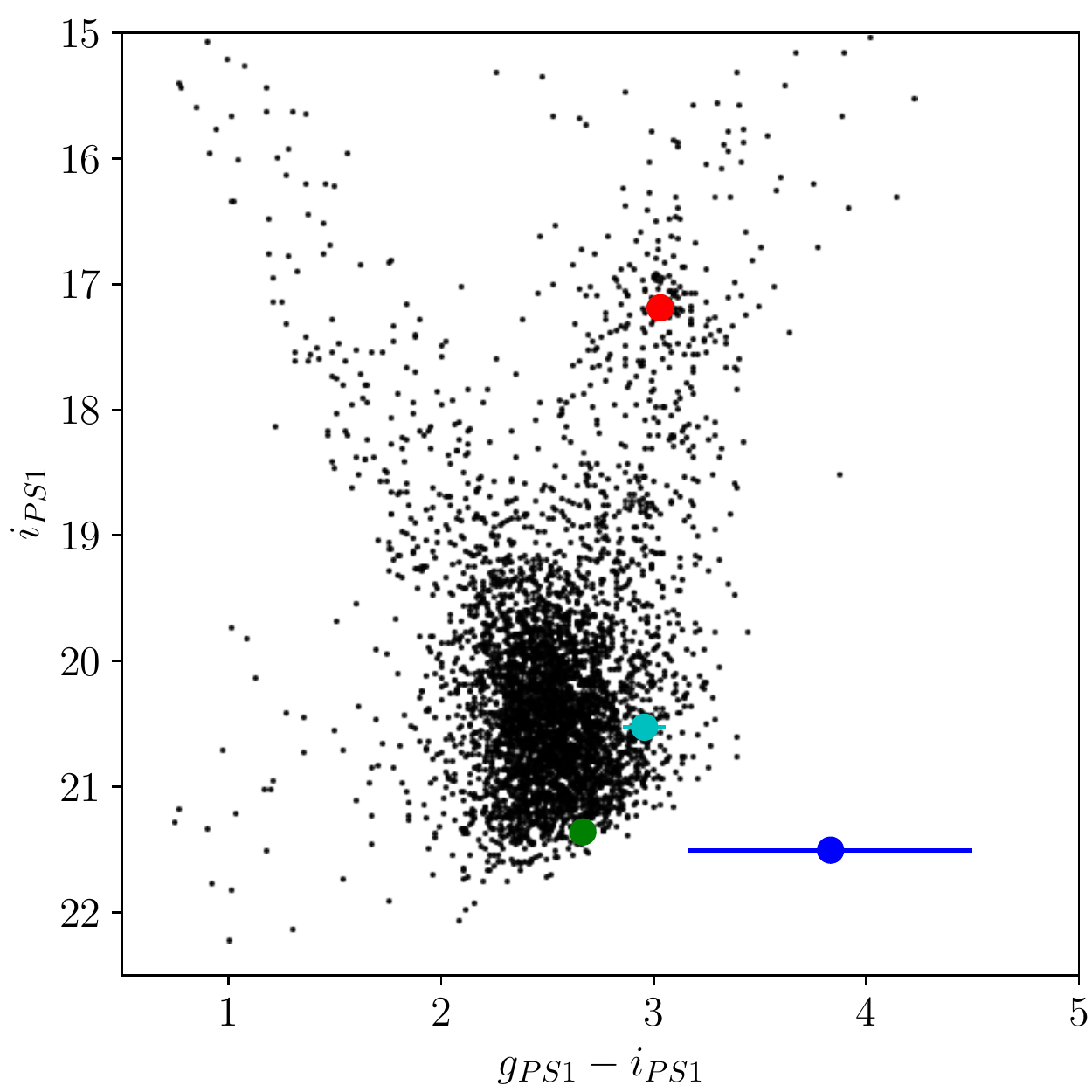}
\caption{Colour-magnitude diagram built from CFHT reference image photometry \citep{Zang18} showing stars within 60" of K2-2016-BLG-0005 (black points) as well as the source star (green circle), red clump position (red circle), and the position of the coincident object in the CFHT reference images (cyan circle). The residual light (i.e., the reference image object minus the light from the source, blue circle) is broadly consistent with a K dwarf at the distance of the lens, but the reference image was constructed using data taken when the source was magnified, so detailed conclusions can not be drawn about whether the residual light is due to the lens.}
\label{fig:cmd}
\end{figure}

To calculate the angular Einstein radius we use the source angular
diameter measured from high-resolution CFHT colour photometry (c.f. right-hand panel of Figure~\ref{fig:k2c9-cfht}). From the model fit to
the CFHT differential photometry light curve we estimated instrumental
source magnitudes and calibrated them to the PanSTARRS-1 system
\citep{Magnier2020} using the \verb+calibrate_flux.py+\footnote{https://github.com/mtpenny/cfht-microlensing} tool provided by \citet{Zang18},
yielding source magnitudes of $g_{\ast}=24.026 \pm 0.013$, $r_{\ast}=22.332 \pm
0.015$, and $i_{\ast}=21.360 \pm 0.015$. We de-reddened these, following the method of
\citet{Yoo2004}, deriving an estimate of the colours and magnitudes of red-clump stars in a $60^{\prime\prime}$ circle around K2-2016-BLG-0005 of $(g-i)=3.03 \pm 0.037$, $(r-i)=1.03 \pm 0.037$ and $i=17.190 \pm 0.007$, respectively; the colour--magnitude diagram used for this is shown in Figure~\ref{fig:cmd}. The intrinsic $V-I$ colour and $I$-band magnitude of the clump
were estimated using \citet{Nataf2016} and transformed
to PanSTARRS magnitudes using the transformations of
\citet{Finkbeiner2016}. We then subtracted these from our measured magnitudes to
yield estimates of the extinction, $A_i=2.39 \pm 0.04$ and reddening
$E(g-i)=2.03 \pm 0.03$ and $E(r-i)=0.799 \pm 0.008$, and dereddened source
colours and magnitudes of {\bf $(g-i)_{\ast,0}=0.64\pm0.03$, $(r-i)_{\ast,0}=0.18\pm0.02$,} $g_{\ast,0}=19.61 \pm 0.05$, $r_{\ast,0}=19.15 \pm 0.05$, and $i_{\ast,0}=18.97 \pm 0.04$. We utilized the colour--surface-brightness relation (CSBR)
for PanSTARRS magnitudes provided by \citet{Zang18} in their equation 7 and Table 3, which were derived from
photometry and relations found by \citet{Boyajian2012, Boyajian2013,
  Boyajian2014}. From these we estimate a source angular diameter of
$\theta_{\ast}=2.12 \pm 0.10~\mu$as using $(g-i)$ and $i$ photometry and relations, and
$\theta_{\ast}=1.79 \pm 0.15~\mu$as using $(r-i)$ photometry and
relations. Combined with the measurement of $\rho$ from the light curve
modelling, we compute the angular Einstein  radius to be
$\theta_E = \theta_{\ast}/\rho = 0.57 \pm 0.03$~mas using the $(g-i)$ source angular
diameter, and $\theta_{\rm E} = 0.48 \pm 0.05$~mas from the $(r-i)$ angular
diameter. Throughout this calculation we estimated uncertainties on each quantity by sampling from Gaussian distributions representing random and systematic uncertainties on the photometric calibration, extinction, reddening, and CSBRs, and combined these with samples from the MCMC chain for the relevant lightcurve parameters; given the proximity of the dereddened source colors to the pivot point of the CSBRs ($g-i=0.58$ and $r-i=0.15$) we adopt the lower end of the CSBR systematic uncertainty range quoted by \citet{Zang18}. Of these uncertainties, the systematic uncertainty on the CSBR and the $i$-band extinction are the largest, and combined dominate the error budget.

The source angular diameters computed from each set of colours are mildly discrepant from one another though are
compatible within 2 sigma. A likely contributor to the
tension may come from an outlier in the photometry of one filter during a
period of high magnification, as these data points are rare, but
provide a large difference in flux over which to measure the source
magnitude. Over the caustic crossing, CFHT gathered
just 4, 4, and 6 data points in the $g$, $r$, and $i$ bands,
respectively. Since the $i$ band photometry is common to both estimates the cause of any discrepancy must arise either from the $g$ or $r$ band. To assess which of the these was
most sensitive to outliers we repeatedly fit for the $g$ and $r$
source flux around peak magnification after successively removing one data point in the time range $BJD-2450000=7485$
to $7575$. The standard deviation between these fits is $0.007$ mag in $g_{\ast}$ and
$0.023$ mag in $r_{\ast}$. We conclude from this that the $r$-band light curve photometry is likely to be the least reliable, so we opt to discard the
$(r-i)$-based estimate of the angular Einstein radius and adopt
the $(g-i)$-based measurement of $\theta_{\rm E}=0.57 \pm 0.03$~mas. The impact of choosing one solution over the other in any case leads to only a $20\%$ variation in the final planet mass.

\subsection{Host and planetary mass}

Combining the microlens parallax, event timescale, and angular Einstein radius we can estimate
the host and planetary masses, lens distance, lens--source relative proper motion, $\mu_{\rm rel}$, and host--planet projected separation, $a_{\perp}$. We find a host mass\footnote{Had we used the $r$-band measurement of $\theta_{\rm E}$, $M_{\rm L}=0.494\pm0.048$ $M_{\odot}$ would have been found.} of $M_{\rm L}=0.58\pm0.04\,{\rm M}_{\odot}$ and a planet mass of $1.10 \pm 0.09\,{\rm M_J}$ at $a_{\perp} = 4.1\pm0.3$~au. Assuming a source distance of $8\pm0.5$~kpc the lens--source relative parallax of $0.068 \pm 0.004$ mas results in a observer--lens distance of $5.2\pm 0.2$~kpc, which favours a planetary system residing in the Galactic disk. We find $\mu_{\rm rel} = 2.7 \pm 0.1$~mas~year$^{-a1}$, which corresponds to a lens--source relative transverse speed of around 70~$\mathrm{km\,s}^{-1}$ at the lens distance, which is comparatively low but not inconsistent with a disk-lens--bulge-source event. The overall results for the lens system are summarised in Table~\ref{table:lens_prop} for both ecliptic degenerate solutions.

\subsection{Planet orbital distance and period} \label{hostsep}

Whilst microlensing can, as in this case, provide a precise measurement of $s$, namely the projected separation between planet and host in units of the Einstein radius, it is not straightforward to translate from $s$ to a deprojected mean orbital radius $a$ and period $P$.

To obtain limits on $a$ and $P$ we performed a Monte-Carlo simulation using the values for $s$, $M_L$, $t_E$, $D_L$ and $\mu_{\rm rel}$ given in Tables~\ref{table:table0} and \ref{table:lens_prop}, together with their errors.  We sample $s$ from a prior distribution given by $dN/d\ln s \propto s^x$, where we take $x = 0.49\pm0.48$ based on \cite{2016ApJ...833..145S} for planets between $0.1 < s < 10$. The error in $x$ is included in our simulation. To convert $s$ to a projected physical separation $a_{\perp}$ we sample the Einstein angular radius $\theta_E = \mu_{\rm rel} t_{\rm E}$ and ${D_L}$ from the observed values and their error distributions to give $a_{\perp} = s \theta_{\rm E} D_{L}$. Assuming a uniform random distribution for both orbital inclination and phase (i.e., circular orbits), we compute $a$ from $a_{\perp}$ and sample $M_L$  to compute $P$. Each sample is assigned a statistical weight according to the probability of the sampled $s$ given its fitted value and error. The final bounds on $a$ and $P$ are calculated separately assuming a logarithmic prior to compute the respective weighted cumulative distribution function.

\begin{figure*}
    \centering
    \includegraphics{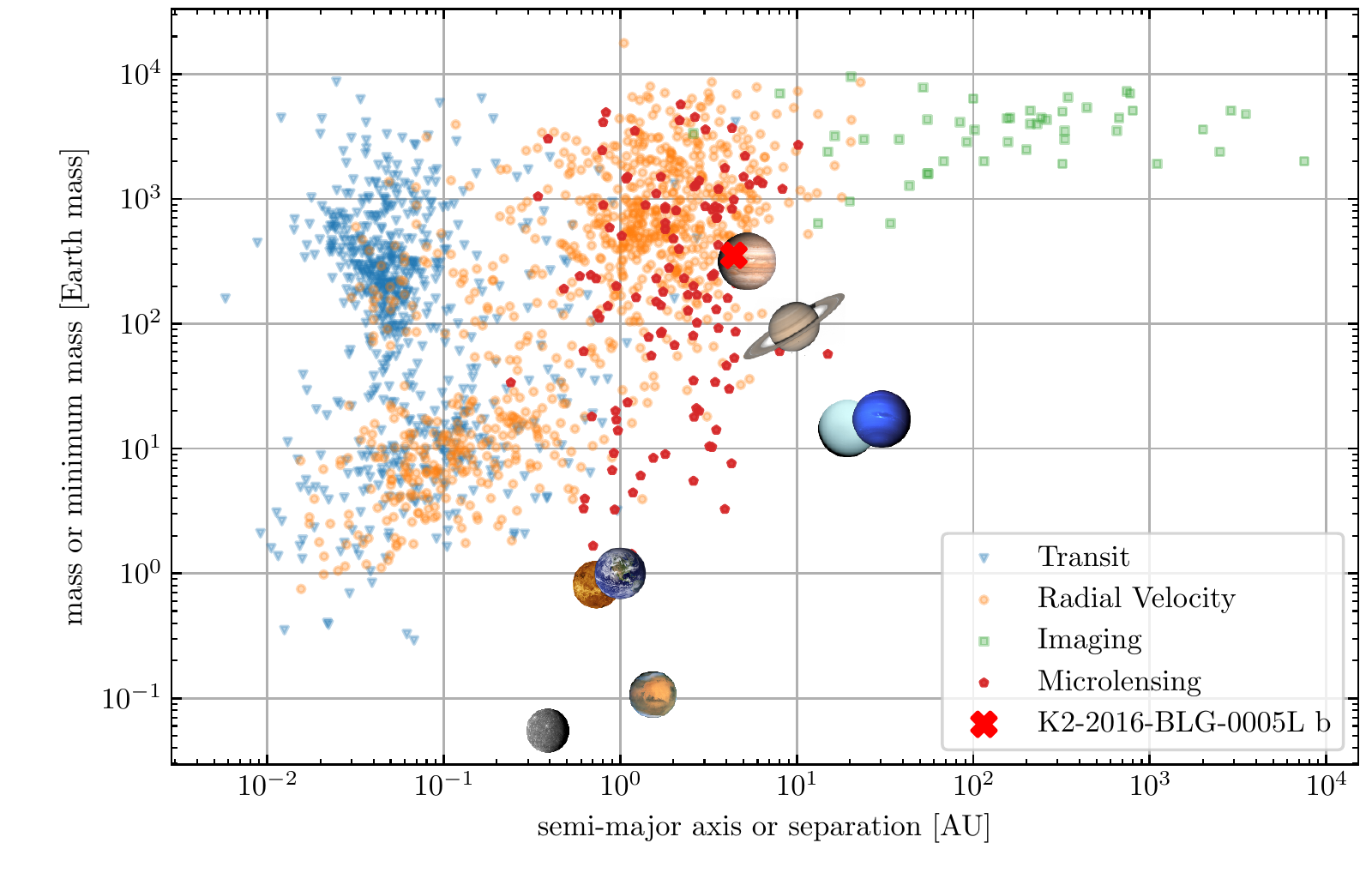}
    \caption{Distribution of mass vs semi-major-axis (or separation) for confirmed exoplanets and Solar System planets (images credit to NASA.). Data are shown on a logarithmic scale, with K2-2016-BLG-0005Lb indicated with a red cross. 
    Exoplanets are indicated using various symbols for different detection techniques and were extracted from the NASA Exoplanet Archive \citep[accessed 2021 November 28;][]{Akeson13}. Exoplanets are shown only if both parameters are provided by the NASA Exoplanet Archive.
    }
    \label{fig:mvr}
\end{figure*}
From this we find that $a = 4.4^{+1.9}_{-0.4}$~au and $P = 13^{+9}_{-2}$~yr, where the central value is the median and the range spans the 68 per cent confidence interval. These values are consistent with K2-2016-BLG-0005Lb being a close analogue of Jupiter (Figure~\ref{fig:mvr}), albeit orbiting a star somewhat smaller than the Sun, such as a mid-K-type dwarf.

We note that the minimum orbital period of at least 11~years is sufficiently long that it is safe to neglect the effects of binary orbital motion on our fits.

\section{Discussion}

We have presented K2-2016-BLG-0005Lb, the first bound exoplanet to be discovered from space-based microlensing observations. It was discovered from dense time series photometry from \emph{K2}C9. Together with simultaneous ground-based data from OGLE, KMTNet, MOA, CFHT and UKIRT we have shown that the data are well modelled by a binary lens system involving a planetary-mass secondary lens orbiting a sub-solar mass host.

The dense \emph{K2}C9 time series provides well resolved magnification caustics on both entry and exit, allowing the Einstein radius to be measured. The combination of spatially well separated simultaneous photometry from the ground and space also enables a precise measurement of the lens--source relative parallax. These measurements allow us to determine a precise planet mass ($1.1\pm 0.1~M_J$), host mass ($0.58\pm0.04~M_{\odot}$) and distance ($5.2\pm 0.2$~kpc).  The inferred host separation of the planet is determined to be $4.4^{+1.9}_{-0.4}$~au and the planet orbital period is $13^{+9}_{-2}$~yr, making this a close analogue of Jupiter orbiting a K-dwarf star. The location of the lens system and its transverse proper motion relative to the background source star ($2.7\pm 0.1$~mas/yr) are consistent with a distant Galactic-disk planetary system microlensing a star in the Galactic bulge.  K2-2016-BLG-0005Lb is more than twice as distant as Kepler-40b, the next furthest exoplanet discovered by \emph{Kepler}. At just 0.6~M$_{\odot}$ its host star is only just above the 0.5~M$_{\odot}$ threshold below which planets as massive as Jupiter are not seen to form within planet formation simulations \citep{2021A&A...656A..72B}.

This discovery was made using a space telescope that was not designed for microlensing observations  and, due to its large pixel size and poor pointing stability, is highly sub-optimal for precision relative photometry towards the highly crowded Galactic Bulge fields. Nonetheless, using the Modified Causal Pixel Method (MCPM), a recently developed photometric method purpose-built to handle K2 microlensing data, we have obtained a direct planet-mass measurement of high precision. The mass measurement precision owes much to the uninterrupted high observing cadence that is facilitated by observing from space.

In 2023, the ESA \emph{Euclid} mission will launch, followed a few years later by the NASA \emph{Roman} mission. Both telescopes will be optimal for exoplanet microlensing discovery towards the Galactic bulge as they will both carry sensitive near-infrared arrays with wide fields, high resolution, and well-characterised point-spread functions. These missions have the capacity to revolutionise our understanding of cool exoplanet demography, a crucial regime for testing theories of planet formation. One of the core science activities of \emph{Roman} will be an exoplanet microlensing survey with 15~min cadence, twice the cadence of \emph{K2}C9. \emph{Roman} will be able to conduct uninterrupted microlensing observations for two 72-day periods per year. An exoplanet microlensing survey is also being considered as an additional science activity for \emph{Euclid}, potentially to be coordinated with that by \emph{Roman} \citep{2022arXiv220209475B}. \emph{Euclid} will be able to observe the bulge for up to 30~days twice per year, though such a campaign would likely only occur towards the end or after the \emph{Euclid} cosmology science program. Both missions will be capable of detecting large numbers of cool, low-mass exoplanets and are expected to be able to make direct mass measurements for a large fraction of events. Since both telescopes will be on halo orbits located at L2, their mutual separation could even provide high precision simultaneous space-based parallax mass measurements to augment those by survey teams on the ground. Their combined data could yield many direct planet mass, orbit and distance measurements using a similar, though in many ways more straightforward, approach to that undertaken in this paper with \emph{K2}C9 data.

\section*{Acknowledgements}
DS acknowledges receipt of a PhD studentship from the UK Science and Technology Facilities Council (STFC). Work by RP was supported by Polish National Agency for Academic Exchange grant ``Polish Returns 2019.'' Work by MTP was partially supported by NASA grants
NNX16AC62G and Louisiana Board of Regents
Support Fund (RCS Award Contract Simple: LEQSF(2020-23)-RD-
A-10). EK acknowledges funding for this work from the STFC (grant ST/P000649/1). Work by B.S.G. was partially supported by the Thomas Jefferson Chair for Space Exploration endowment from the Ohio State University. J.C.Y. acknowledges support from N.S.F Grant No. AST-2108414.
Y.S. acknowledges support from BSF Grant No. 2020740.

This paper includes data collected by the Kepler mission and obtained from the Mikulski Archive for Space Telescopes (MAST) data archive at the Space Telescope Science Institute (STScI). Funding for the Kepler mission is provided by the NASA Science Mission directorate. 

This research uses data obtained through the Telescope Access Program (TAP), which has been funded by the National Astronomical Observatories of China, the Chinese Academy of Sciences (the Strategic Priority Research Program "The Emergence of Cosmological Structures" Grant No. XDB09000000), and the Special Fund for Astronomy from the Ministry of Finance. Based on observations obtained with MegaPrime/MegaCam, a joint project of CFHT and CEA/DAPNIA, at the Canada-France-Hawaii Telescope (CFHT) which is operated by the National Research Council (NRC) of Canada, the Institut National des Science de l'Univers of the Centre National de la Recherche Scientifique (CNRS) of France, and the University of Hawaii.

This research has made use of the KMTNet system operated by the Korea Astronomy and Space Science Institute (KASI) and the data were obtained at three host sites of CTIO in Chile, SAAO in South Africa, and SSO in Australia.

The MOA project is supported by JSPS KAK-ENHI Grant Number JSPS24253004, JSPS26247023, JSPS23340064, JSPS15H00781, JP16H06287,17H02871 and 19KK0082.

The OGLE project has received funding from the National Science Centre, Poland, grant MAESTRO 2014/14/A/ST9/00121 to A.U. 

UKIRT is currently owned by the University of Hawaii (UH) and operated by the UH Institute for Astronomy; operations are enabled through the cooperation of the East Asian Observatory. When the 2016 data reported here were acquired, UKIRT was supported by NASA and operated under an agreement among the University of Hawaii, the University of Arizona, and Lockheed Martin Advanced Technology Center; operations were enabled through the cooperation of the East Asian Observatory. We furthermore acknowledge the support from NASA HQ for the UKIRT observations in connection with K2C9.

This research has made use of "Aladin sky atlas" developed at CDS, Strasbourg Observatory, France.

\section*{Data Availability}

Data from the K2C9 campaign can be retrieved from the Mikulski Archive for Space Telescopes at \url{https://archive.stsci.edu/k2/}.



\bibliographystyle{mnras}



\input{affils}

\appendix

\section{Binary parameterisation}\label{section:appendix}

The following equations were used to convert from the planetary parameterisation ($t_{0,\rm pl}$,$u_{0,\rm pl}$,$t_{\rm E,\rm pl}$) to the conventional binary parameterisation ($s$,$q$,$\alpha$). The parameters $\gamma$, $u$, $q$ and $\tau$, defined in Eqn~\ref{eqn:wide} are used throughout this section.

For the wide topology models we use the following transformations:
\begin{gather}
    \gamma = \frac{t_{\rm E, \rm pl}}{t_{\rm E}}, \nonumber \\
    u = u_0 + \gamma u_{0,\rm pl}, \nonumber \\
    \tau = \frac{t_{0,\rm pl} - t_0}{t_{\rm E}}, \nonumber \\
    u^{\prime} = \sqrt{u^2 + \tau^2}, \nonumber \\
    s = \frac{1}{2} \Big(u^{\prime} + \sqrt{u^{\prime2} + 4} \Big) \nonumber \\ 
    q = \gamma^2 \nonumber \\
    \alpha = 2\pi - \arcsin{\bigg(\frac{u}{u^{\prime}}\bigg)},
    \label{eqn:wide}
\end{gather}
where the angle $\alpha$ is in radians.

For the close topology models we use the following transformations:
\begin{gather}
    \delta = \big(u^{\prime2} + 2\big) + 4 \big(4\gamma^2+1\big)\big(u^{\prime2}-1\big), \nonumber \\
    \beta = \arctan{\bigg(\frac{\tau}{u_0}\bigg)}, \nonumber \\
    s = \sqrt{-\frac{u^{\prime2}+2\sqrt{\delta}}{2\big(u^{\prime2}-1\big)}}, \nonumber \\
    \eta = \frac{2\gamma}{s\sqrt{1+s^2}}, \nonumber \\
    \theta = \arctan{\bigg(\frac{\eta}{s^{-1}-s}\bigg)}, \nonumber \\
    \alpha = (\theta+\beta) + \pi/2
    \label{eqn:close}
\end{gather}

For the resonant case, the value of $u_0$ must be constrained in order to ensure the second caustic approach, by reassigning it the value $u_{0,\rm new}$. The parameter $\eta$ is also reused from Eqn~\ref{eqn:close}:
\begin{gather}
    \nu = 16\frac{q}{\tau^2 + (u_{0,\rm pl} - u_0)}, \nonumber \\
    s = \sqrt{\frac{\sqrt{4\nu+1}-1}{2}}, \nonumber \\
    \alpha = \pi/2 - \arctan{\bigg(\frac{u_{0,\rm pl} - u_0}{\tau}\bigg)}, \nonumber \\
    u_{0,\rm new} = \bigg(s + \frac{1}{s} - \frac{\eta}{\tan{(\pi - \alpha)}}\bigg)
      \sin{(\pi - \alpha)} - u_{0,\rm pl}
\end{gather}

\begin{figure*}
    \centering
    \includegraphics[width=\textwidth]{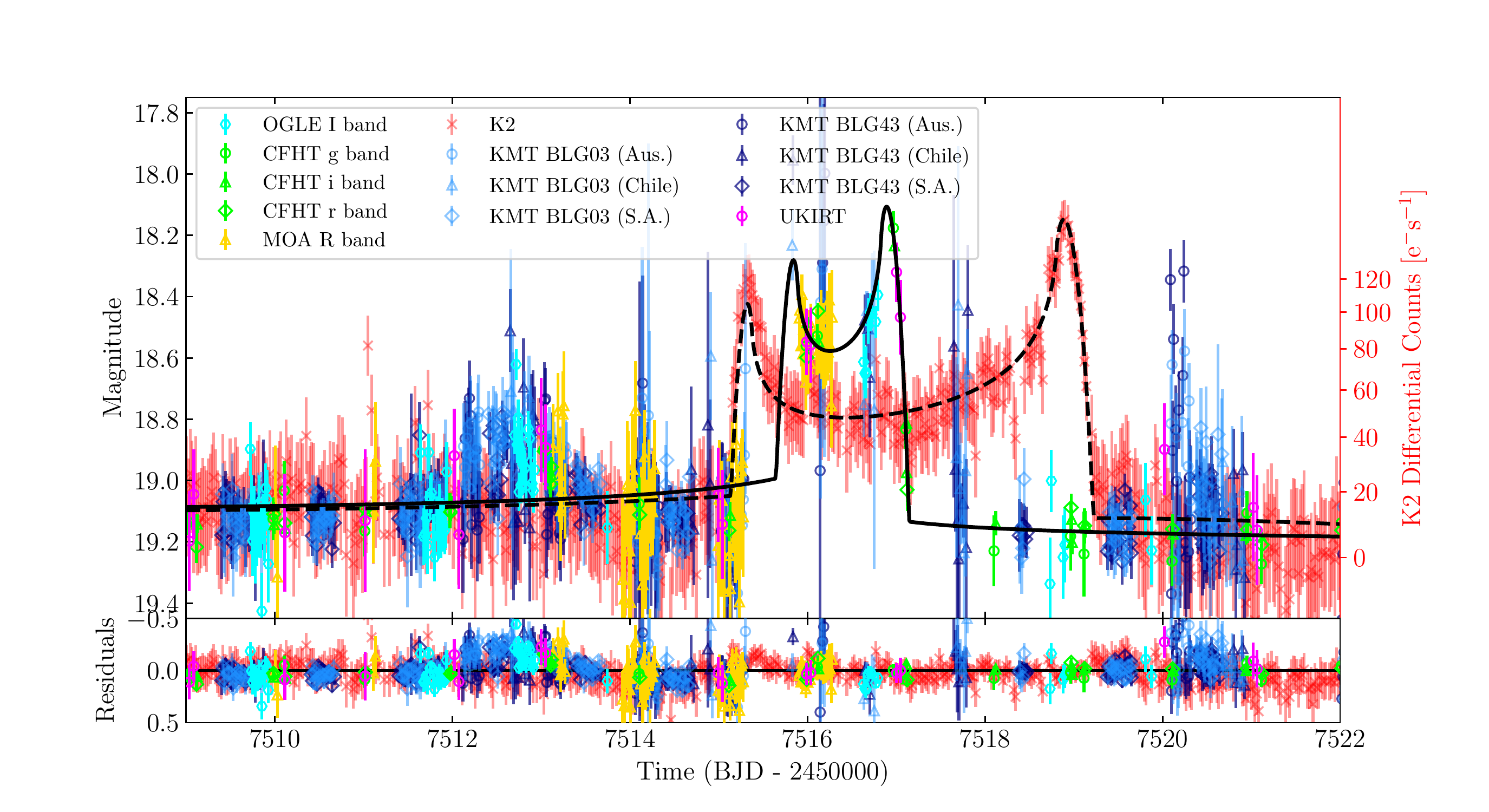}
    \caption{Same as Figure~\ref{fig:wide} but showing the best-fit close topology solution. Note the failure of this solution to account for the pre-caustic peak at around BJD$-2450000 = 7512.5$~days that is seen in multiple ground-based survey data. This fit has $\Delta \chi^2 = \chi^2_{\rm close} - \chi^2_{\rm wide}$ = 693.}
    \label{fig:close}
\end{figure*}

\begin{figure*}
    \centering
    \includegraphics[width=\textwidth]{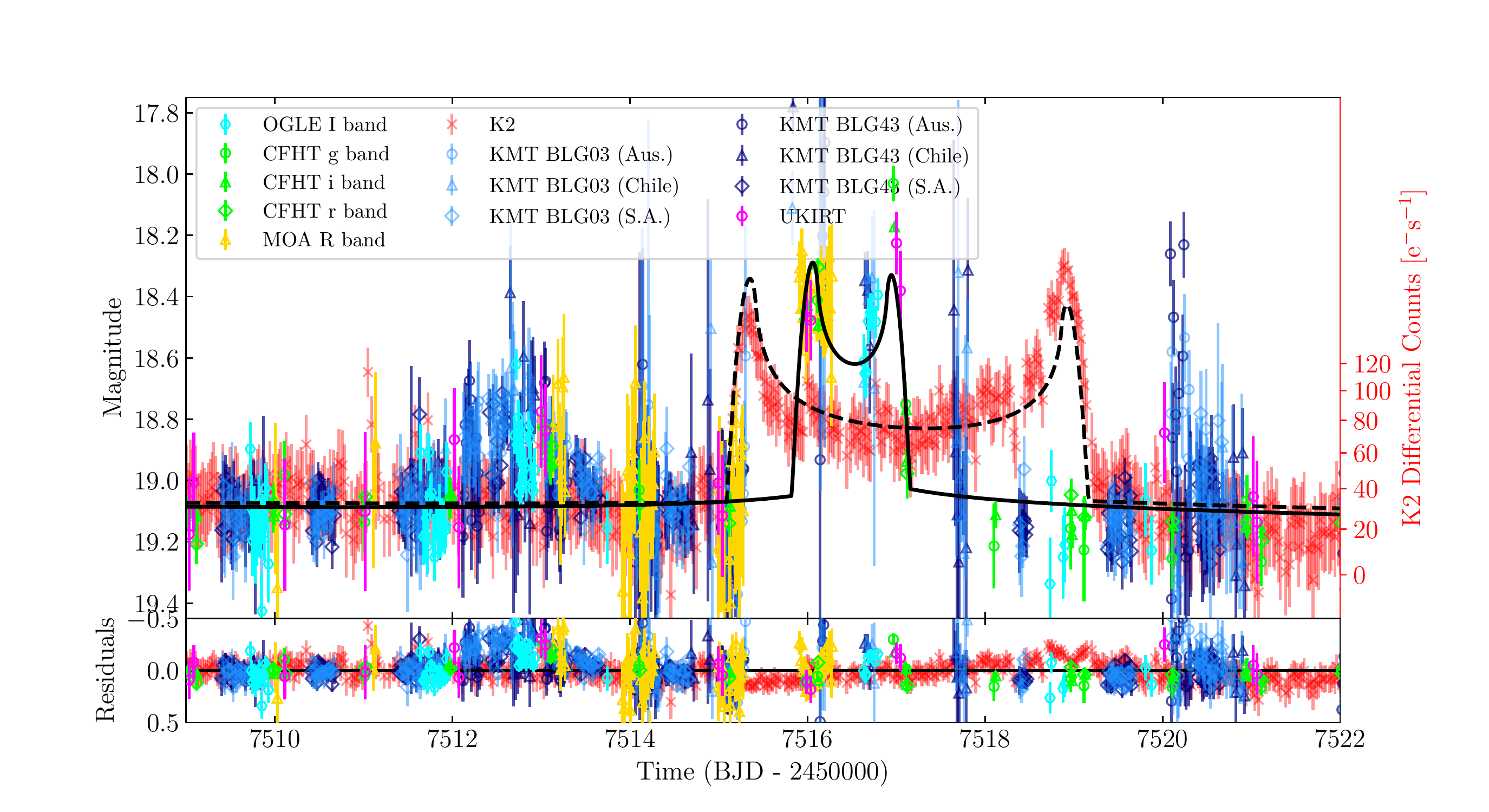}
    \caption{Same as Figure~\ref{fig:wide} but showing the best-fit resonant topology solution. Note the failure of this solution both to describe adequately the caustic behaviour seen in the \emph{K2}C9 data and to account for the pre-caustic peak at around BJD$-2450000 = 7512.5$~days that is seen in multiple ground-based survey data. This fit has $\Delta \chi^2 = \chi^2_{\rm resonant} - \chi^2_{\rm wide}$ = 1,308.}
    \label{fig:resonant}
\end{figure*}

\begin{figure*}
    \centering
    \includegraphics[width=\textwidth]{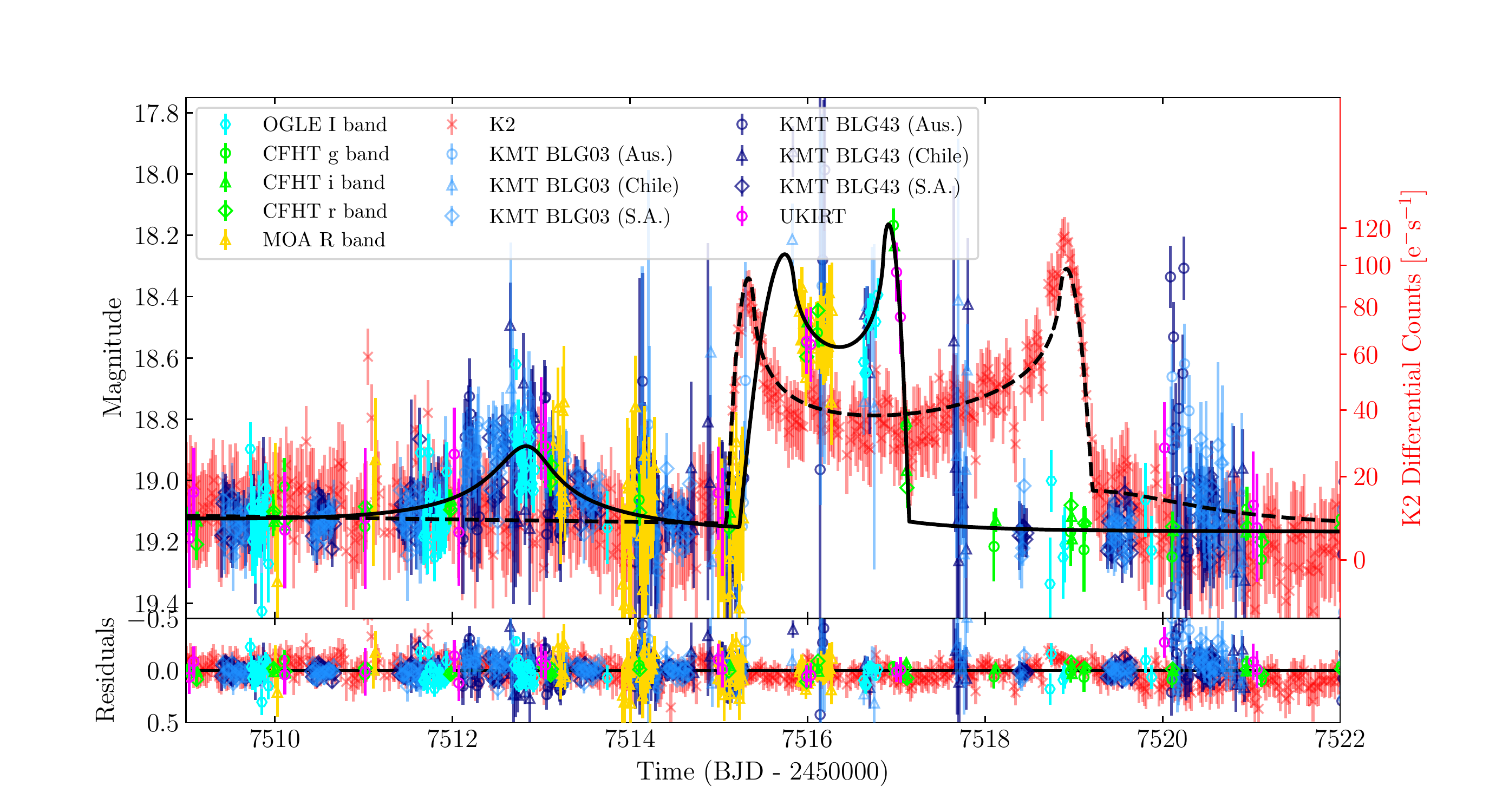}
    \caption{Same as Figure~\ref{fig:wide} but showing the ecliptic degenerate solution, acquired by transforming $u_0 \rightarrow -u_0$, $\alpha \rightarrow -\alpha$ and $\pi_{\rm E,N} \rightarrow -\pi_{\rm E,N}$. Although performing better than the close and resonant solutions and allowing for the cusp approach at BJD$-2450000 = 7512.5$~days, the model's caustic exit shows a noticeable underestimation in flux, contributing to the marginally inferior $\chi^2$.}
    \label{fig:ecl_sol}
\end{figure*}

\begin{figure*}
    \centering
    \includegraphics[width=15cm]{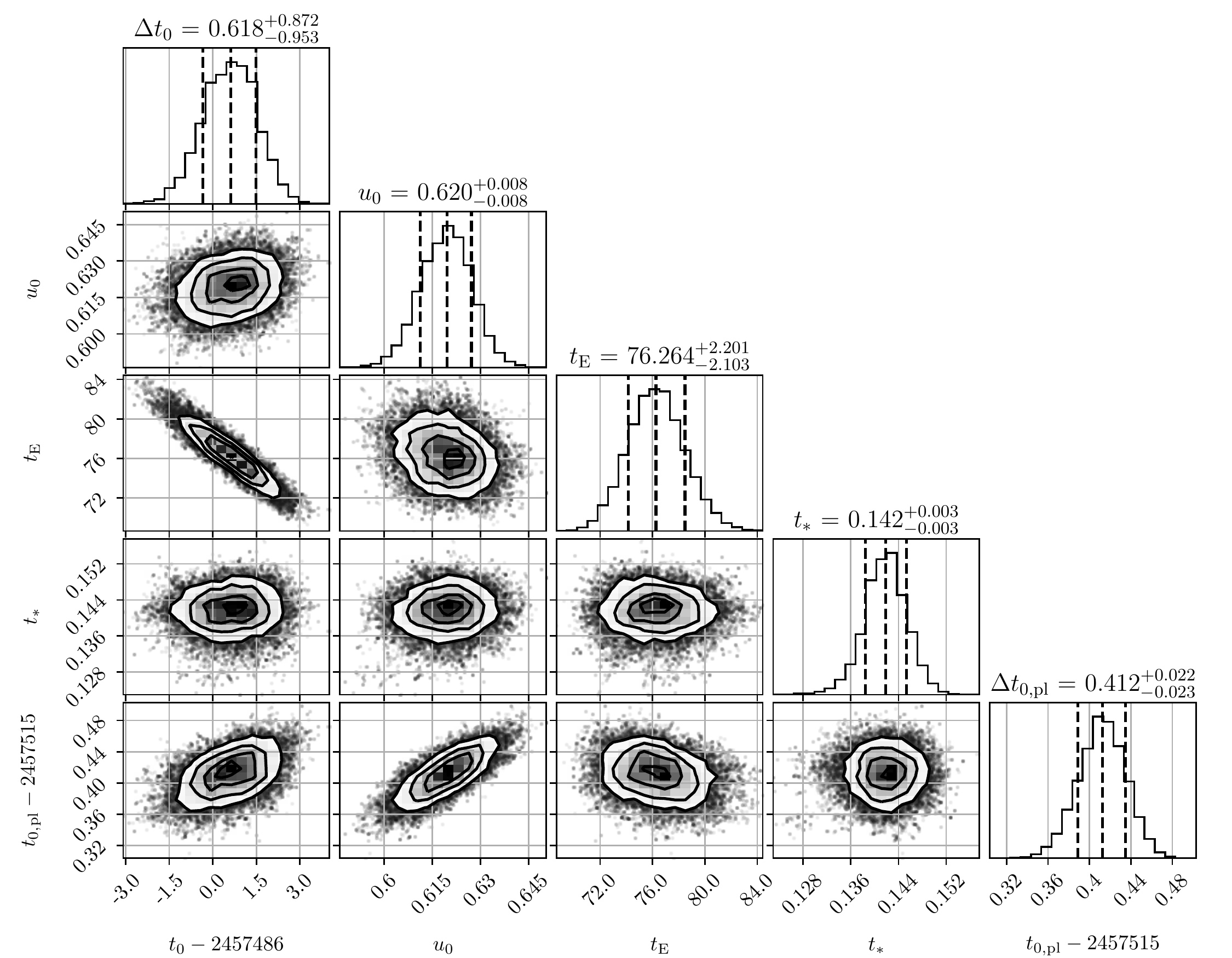}
    \caption{Corner plot posterior distributions for the wide topology best-fit solution. This part of the figure is the first of three sections spanning the full corner plot; the following figure parts show the remaining sections. The corner plots in this paper use code from \citet{corner}.}
    \label{fig:cornerTop}
\end{figure*}

\begin{figure*}
    \centering
    \includegraphics[width=13.5cm]{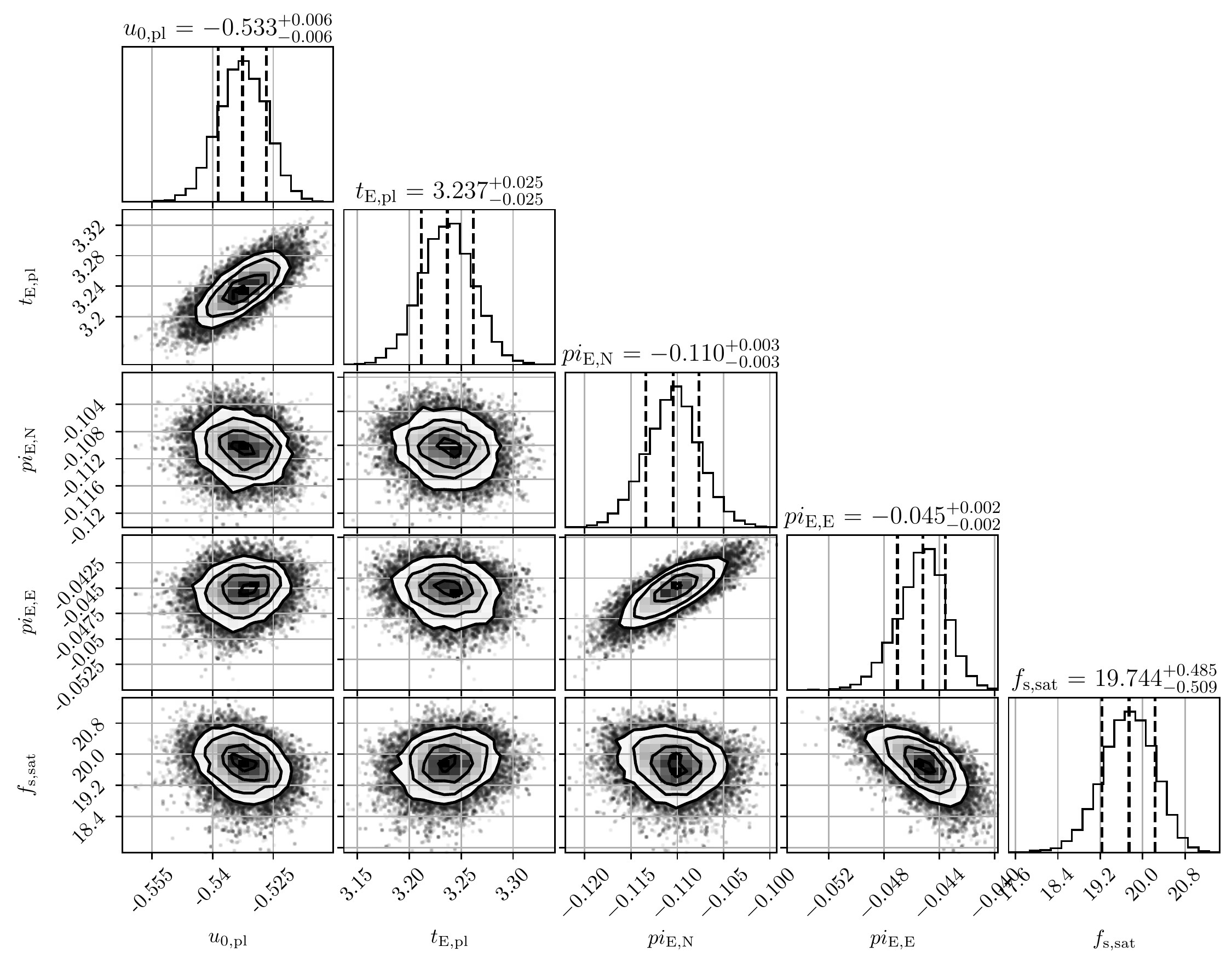}
    \contcaption{The second of three sections of the corner plot posterior distributions for the wide topology best-fit solution.}
\end{figure*}

\begin{figure*}
    \centering
    \includegraphics[width=13.5cm]{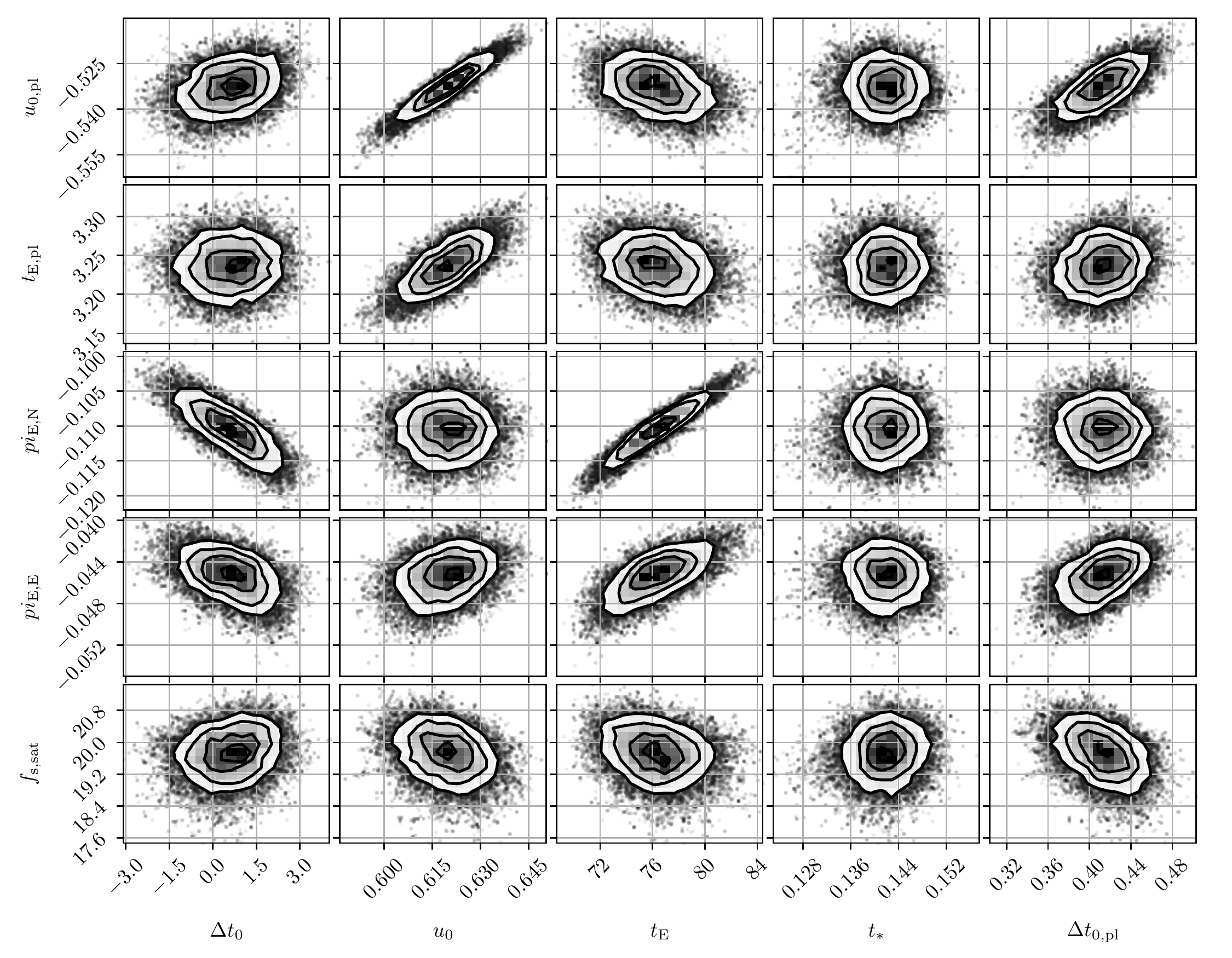}
    \contcaption{The last of three sections of the corner plot posterior distributions for the wide topology best-fit solution.}
\end{figure*}


\bsp	
\label{lastpage}
\end{document}

%% file: authors.tex
\author[D. Specht et al.]
{D.~Specht,$^{1}$
R.~Poleski,$^{2,O}$
M.T.~Penny,$^{3,C}$
E.~Kerins,$^{1}$\thanks{Corresponding author: Eamonn.Kerins@manchester.ac.uk} 
I.~McDonald,$^{1,4}$ 
Chung-Uk~Lee,$^{5,6,K}$
A.~Udalski,$^{2,O}$ \newauthor
I.A.~Bond,$^{7,M}$
Y.~Shvartzvald,$^{8,U}$
Weicheng~Zang,$^{9,K,C}$
R.A.~Street,$^{10,S}$
D.W.~Hogg,$^{11,12,S}$
B.S.~Gaudi,$^{13,S,U}$\newauthor
T.~Barclay,$^{14,S}$
G.~Barentsen,$^{15,S}$ 
S.B.~Howell,$^{15,S}$
F.~Mullally,$^{15,S}$
C.B.~Henderson,$^{16,S,U}$
S.T.~Bryson,$^{15,E}$\newauthor
D.A.~Caldwell,$^{15,E}$
M.R.~Haas,$^{15,E}$
J.E.~Van~Cleve,$^{15,E}$
K.~Larson,$^{17,E}$
K.~McCalmont,$^{17,E}$
C.~Peterson,$^{17,E}$\newauthor 
D.~Putnam,$^{17,E}$
S.~Ross,$^{17,E}$
M.~Packard,$^{18,E}$
L.~Reedy,$^{18,E}$
Michael~D.~Albrow,$^{19,K}$ 
Sun-Ju~Chung,$^{5,6,K}$\newauthor 
Youn~Kil~Jung,$^{5,K}$
Andrew~Gould,$^{20,13,K}$  
Cheongho~Han,$^{21,K}$ 
Kyu-Ha~Hwang,$^{5,K}$
Yoon-Hyun~Ryu,$^{5,K}$\newauthor 
In-Gu~Shin,$^{5,K}$
Hongjing~Yang,$^{9a,K}$
Jennifer~C.~Yee,$^{22,K}$
Sang-Mok~Cha,$^{5,23,K}$ 
Dong-Jin~Kim,$^{5,K}$\newauthor
Seung-Lee~Kim,$^{5,6,K}$
Dong-Joo~Lee,$^{5,K}$ 
Yongseok~Lee,$^{5,23,K}$ 
Byeong-Gon~Park,$^{5,6,K}$ 
Richard~W.~Pogge,$^{13,K}$\newauthor
M.K.~Szyma\'nski,$^{2,O}$
I.~Soszy\'nski,$^{2,O}$
K.~Ulaczyk,$^{24,2,O}$
P.~Pietrukowicz,$^{2,O}$
Sz.~Koz{\l}owski,$^{2,O}$\newauthor
J.~Skowron,$^{2,O}$
P.~Mr\'oz,$^{2,O}$
Shude~Mao,$^{25,26,C}$ 
Pascal~Fouqu\'e$^{27,28,C}$ 
Wei~Zhu$^{25,C}$ 
F.~Abe,$^{29,M}$ 
R.~Barry,$^{30,M}$\newauthor
D.P.~Bennett,$^{30,31,M}$ 
A.~Bhattacharya,$^{30,31,M}$
A.~Fukui,$^{32,33,M}$
H.~Fujii,$^{29,M}$ 
Y.~Hirao,$^{34,M}$ 
Y.~Itow,$^{29,M}$\newauthor
R.~Kirikawa,$^{34,M}$
I.~Kondo,$^{34,M}$ 
N.~Koshimoto,$^{35,M}$ 
Y.~Matsubara,$^{29,M}$
S.~Matsumoto,$^{34,M}$ 
S.~Miyazaki,$^{34,M}$\newauthor
Y.~Muraki,$^{29,M}$ 
G.~Olmschenk,$^{30,M}$
C.~Ranc,$^{36,M}$ 
A.~Okamura,$^{34,M}$ 
N.J.~Rattenbury,$^{37,M}$ 
Y.~Satoh,$^{34,M}$ \newauthor
T.~Sumi,$^{34,M}$ 
D.~Suzuki,$^{34,M}$ 
S.I.~Silva,$^{38,30,M}$
T.~Toda,$^{34,M}$
P.J.~Tristram,$^{39,M}$ 
A.~Vandorou,$^{30,31,M}$
H.~Yama,$^{34}$ \newauthor
C.~Beichman,$^{16,U}$
G.~Bryden,$^{40,U}$ S.~Calchi~Novati$^{16,U}$\newauthor
\\
Affiliations are listed at the end of the paper
}

%% file: affils.tex
\section*{Author affiliations}
$^{1}$Jodrell Bank Centre for Astrophysics, Department of Physics and Astronomy, University of Manchester, Oxford Road, Manchester M13 9PL, UK\\
$^{2}$Astronomical Observatory, University of Warsaw, Al. Ujazdowskie 4, 00-478 Warsaw, Poland\\
$^{3}$Louisiana State University, 261-B Nicholson Hall, Tower Dr., Baton Rouge, LA 70803-4001, USA\\
$^{4}$Department of Physical Sciences, The Open University, Walton
Hall, Milton Keynes, MK7 6AA, UK\\
$^{5}$Korea Astronomy and Space Science Institute, Daejon 34055, Republic of Korea\\
$^{6}$Korea University of Science and Technology, Korea, 217 $^{7}$Institute for Natural and Mathematical Sciences, Massey University, Private Bag 102904 North Shore Mail Centre, Auckland 0745, New Zealand\\
$^{8}$Department of Particle Physics and Astrophysics, Weizmann Institute of Science, Rehovot 76100, Israel\\
$^{9}$ Department of Astronomy and Tsinghua Centre for Astrophysics, Tsinghua University, Beijing 100084, China\\
$^{10}$Las Cumbres Observatory Global Telescope Network, 6740 Cortona Drive, Suite 102, Goleta, CA 93117, USA\\
$^{11}$Center for Cosmology and Particle Physics, Department of Physics, New York University, 4 Washington Place, Room 424, New York, NY 10003, USA \\
$^{12}$Center for Data Science, New York University, 726 Broadway, 7th Floor, New York, NY 10003, USA \\
$^{13}$Department of Astronomy, The Ohio State University, 140 West 18th Avenue, Columbus, OH 43210, USA\\
$^{14}$ University of Maryland, Baltimore County, 1000 Hilltop Circle, Baltimore, MD 21250, USA \\
$^{15}$NASA Ames Research Center, Moffett Field, CA 94035, USA \\
$^{16}$IPAC, Mail Code 100-22, Caltech, 1200 E. California Blvd., Pasadena, CA 91125, USA\\
$^{17}$Ball Aerospace \& Technologies, Boulder, CO 80301, USA \\
$^{18}$Laboratory for Atmospheric and Space Physics, University of Colorado at Boulder, Boulder, CO 80303, USA \\
$^{19}$University of Canterbury, Department of Physics and Astronomy, Private Bag 4800, Christchurch 8020, New Zealand\\
Gajeong-ro Yuseong-gu, Daejeon 34113, Korea\\
$^{20}$Max-Planck-Institute for Astronomy, K$\rm \ddot{o}$nigstuhl 17, 69117 Heidelberg, Germany\\
$^{21}$Department of Physics, Chungbuk National University, Cheongju 28644, Republic of Korea\\
$^{22}$Center for Astrophysics $|$ Harvard \& Smithsonian, 60 Garden St., Cambridge, MA 02138, USA\\
$^{23}$School of Space Research, Kyung Hee University, Yongin 17104, Republic of Korea\\
$^{24}$Department of Physics, University of Warwick, Coventry CV4 7AL, UK\\
$^{25}$ Physics Department and Tsinghua Centre for Astrophysics, Tsinghua University, Beijing 100084, China\\
$^{26}$ National Astronomical Observatories, Chinese Academy of Sciences, A20 Datun Rd., Chaoyang District, Beijing 100012, China\\
$^{27}$ CFHT Corporation, 65-1238 Mamalahoa Hwy, Kamuela, Hawaii 96743, USA\\
$^{28}$ Universit\'e de Toulouse, UPS-OMP, IRAP, Toulouse, France\\
$^{29}$Institute for Space-Earth Environmental Research, Nagoya University, Nagoya 464-8601, Japan\\
$^{30}$Code 667, NASA Goddard Space Flight Center, Greenbelt, MD 20771, USA\\
$^{31}$Department of Astronomy, University of Maryland, College Park, MD 20742, USA\\
$^{32}$Department of Earth and Planetary Science, Graduate School of Science, The University of Tokyo, 7-3-1 Hongo, Bunkyo-ku, Tokyo 113-0033, Japan\\
$^{33}$Instituto de Astrof\'isica de Canarias, V\'ia L\'actea s/n, E-38205 La Laguna, Tenerife, Spain\\
$^{34}$Department of Earth and Space Science, Graduate School of Science, Osaka University, Toyonaka, Osaka 560-0043, Japan\\
$^{35}$Department of Astronomy, Graduate School of Science, The University of Tokyo, 7-3-1 Hongo, Bunkyo-ku, Tokyo 113-0033, Japan\\
$^{36}$Sorbonne Universit\'e, CNRS, UMR 7095, Institut d'Astrophysique de Paris, 98 bis bd Arago, 75014 Paris, France\\
$^{37}$Department of Physics, University of Auckland, Private Bag 92019, Auckland, New Zealand\\
$^{38}$Department of Physics, The Catholic University of America, Washington, DC 20064, USA\\
$^{39}$University of Canterbury Mt.\ John Observatory, P.O. Box 56, Lake Tekapo 8770, New Zealand\\
$^{40}$ Jet Propulsion Laboratory, California Institute of Technology, 4800 Oak Grove Drive, Pasadena, CA 91109, USA \\
\\
$^{C}$ K2C9-CFHT Multi-Color Microlensing Survey \\
$^E$ K2C9 Engineering Team\\
$^K$ Korean Microlensing Telescope Network (KMTNet) \\
$^M$ Microlensing Observations in Astrophysics (MOA) \\
$^O$ Optical Gravitational Lensing Experiment (OGLE) \\
$^S$ K2C9 Science Team \\
$^U$ UKIRT Microlensing Survey
https://www.overleaf.com/project/5f61d714c718570001ba52c5